\documentclass[aip,amsmath,amssymb,floatfix, reprint]{revtex4-1}
\usepackage{dcolumn}
\usepackage{bm}%
\usepackage{enumerate}%
\usepackage{longtable}
\usepackage{color}
\usepackage[utf8]{inputenc}

\expandafter\ifx\csname package@font\endcsname\relax\else
 \expandafter\expandafter
 \expandafter\usepackage
 \expandafter\expandafter
 \expandafter{\csname package@font\endcsname}%
\fi

\usepackage{graphicx}
\usepackage{epstopdf}
\usepackage{morefloats}
\begin{document}

\title{Cross Sections for Electron Collisions with NF$_3$ } 
%
\author{ Mi-Young Song$^{a)}$}                
\affiliation{Plasma Technology Research Center, National Fusion Research
Institute, 814-2, Osikdo-dong, Gunsan, Jeollabuk-do, 573-540, South Korea}
\email{corresponding author at mysong@nfri.re.kr}
\author{Jung-Sik Yoon}                
\affiliation{Plasma Technology Research Center, National Fusion Research
Institute, 814-2, Osikdo-dong, Gunsan, Jeollabuk-do, 573-540, South Korea}
\author{Hyuck Cho }                   
\affiliation{Department of Physics, Chungnam National University, Daejeon
305-764, South Korea}
\author{Grzegorz P. Karwasz }                   
\affiliation{Faculty of Physics, Astronomy and Applied Informatics, University Nicolaus Copernicus, Grudziadzka 5, 87-100, Toru\'n, Poland}
\author{Viatcheslav Kokoouline }                   
\affiliation{Department of Physics, University of Central Florida, Orlando, FL 32816, USA}
\author{Yoshiharu Nakamura }                   
\affiliation{6-1-5-201 Miyazaki, Miyamae, Kawasaki, 216-0033, Japan}
\author{James R. Hamilton }                   
\affiliation{Department of Physics and Astronomy, University College London,
Gower Street, London WC1E~6BT, UK}
\author{Jonathan Tennyson }                   
\affiliation{Department of Physics and Astronomy, University College London,
Gower Street, London WC1E~6BT, UK}

\revised{\today}%
\begin{abstract}
Cross section data are compiled from the literature for electron
collisions with  nitrogen trifluoride  (NF$_3$) molecules. Cross sections are
collected and reviewed for total scattering, elastic scattering, momentum
transfer, excitations of rotational and vibrational states, dissociation,
ionization, and dissociative attachment.  For each of these
processes, the recommended values of the cross sections are presented. The
literature has been surveyed up to the end of 2016.
\end{abstract} 
\maketitle


\section{\label{sec:int}Introduction}
Nitrogen trifluoride or trifluoramine (NF$_3$) gas is widely used in plasma
processing technology. 
NF$_3$ is used in a number of plasma processes where it is often used as a source 
of F atoms due to ease of production these atoms 
via dissociative electron attachment (DEA) and electron impact dissociation both
from NF$_3$ itself and from NF$_2$ and NF fragment species.
The exothermicity from these  dissociative 
processes also provides an important gas heating mechanism.
Use of NF$_3$ in plasma etching, particularly in
mixtures with O$_2$, see Ref. \cite{huang2017insights}, provides 
a source of F$^-$ ions due to enhanced dissociative electron attachment
process at low (about 1 eV) energies. 
NF$_3$ is widely used 
for semiconductor fabrication processes which include direct etching
\cite{pkc00,vec00}, reactor cleaning \cite{94bcc} and remote plasma
sources \cite{kas98}, where use of pure NF$_3$ typically
limits the reactants reaching the processing chamber to
F$_x$ and NF$_x$ species only.  NF$_3$ is also used in the production of thin films \cite{goh90,kur02}
and solar cells \cite{park09,gao13}; it provides the initial gas for the HF chemical laser \cite{Lin71,raz16,pos16}.
NF$_3$ is actually a greenhouse gas with a very high global warming potential which
has led to concern on how it is used in the various technologies discussed above \cite{yang16}.
In spite of its importance, experimental studies of electron scattering on NF$_3$ are rather sparse: 
for total\cite{szmytkowski2004nf3} and elastic\cite{boesten1996vibrationally}
cross section measurements come from single laboratories, more measurements
exist for ionization\cite{tarnovsky1994electron,haaland2001,rahman2012electron} and
dissociative electron
attachment\cite{Harland1974nf3,nandi2001nf3,chantry1979dissociative}. 
In absence of experiments, several
calculations\cite{rescigno1995low,joucoski2002elastic,goswami2013cross,
hamilton2017nf3} have been performed. Some reference cross sections based both
on experiments and calculations were reported by  Lisovskiy {\it et al.}
\cite{lisovskiy2014electron} in modeling electron transport coefficients and by
Huang {\it et al.}  \cite{huang2017insights} for modeling remote
plasma sources in NF$_3$ mixtures. Here we perform a detailed analysis of
available data for electron scattering on NF$_3$, to yield recommended total,
elastic, momentum transfer, ionization, dissociation into neutrals, and
vibrational, rotational, and electronic excitation cross sections.  
In the ground electronic state $^1A'$ the  molecule has a shape of a pyramid of
the $C_{3v}$ group with fluorine atoms forming an equilateral triangle. Due to
its symmetry, the dipole moment of the molecule is aligned along the $C_3$
symmetry axis. Geometry, electric dipole moment, and rotational constants are
specified in Table \ref{tab:properties}. 
\begin{table}[!hbp]
\caption{\label{tab:properties} Properties of  NF$_3$ at the equilibrium position of the ground electric state. $A$, $B$, and $C$ are rotational constants;
$\alpha_0$ is the spherical dipole polarizability.}
\begin{tabular}{cc}
\hline\hline
Property & Value \\
\hline
F-N bond length \cite{nist2015nf3}  & 1.365\AA \\
FNF angle \cite{nist2015nf3}& 102.4$^\circ$ \\
Dipole moment \cite{nist2015nf3} & 0.235 D \\
$A=B$ \cite{novick1996hyperfine} & 10.6810819(15)GHz  \\
$C$ \cite{novick1996hyperfine} & 5.8440 GHz \\
$\alpha_0$ \cite{crc2003nf3}& $3.62\times 10^{-30}$ m$^3$ \\
\hline\hline
\end{tabular}
\end{table}


\section{\label{sec:tcs}Total Scattering Cross Section}
\begin{figure}[!tbp]
\includegraphics [width=8cm]{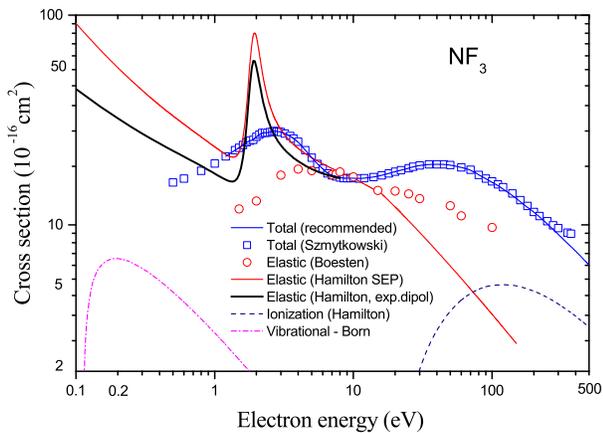}                                
\caption{\label{fig:tot01} Total cross sections by Szmytkowski {\it et al.}
\cite{szmytkowski2004nf3} compared to experimental integral elastic cross
sections of Boesten {\it et al.}\cite{boesten1996vibrationally}, integral
vibrational excitation (Born approximation for the $\nu_3$ IR active mode) and total
ionization (theory by Rahman {\it et al.} \cite{rahman2012electron}). }
\end{figure}

Practically, absolute data by Szmytkowski {\it et al.}\cite{szmytkowski2004nf3}
at 0.5-370 eV collision energy is the only measurement of total cross section in
NF$_3$. The beam attenuation deBeer-Lambert’s method was used, with a 3 cm long
scattering cell and $2 \times 10^{-3} sr$ mean angular resolution. Systematic
errors declared (gas outflow from scattering cell, determination of the
scattering length, current non-linearity, pressure and temperature measurements)
are within 5\%, out of which the declared angular resolution error is 0.2\% at
low energies, rising to 1\% at 100 eV and 2-3\% in the high energy limit. The
statistical spread (one standard deviation of their weighted mean values) is
1.5\% below 1 eV and below 1\% at intermediate energies. 
Total cross sections \cite{szmytkowski2004nf3} are compared to experimental
elastic \cite{boesten1996vibrationally}, ionization \cite{rahman2012electron}
and vibrational excitation (calculated in Born approximation) in Fig.
\ref{fig:tot01}. 

\begin{figure}[!hbp]
\includegraphics [width=8cm]{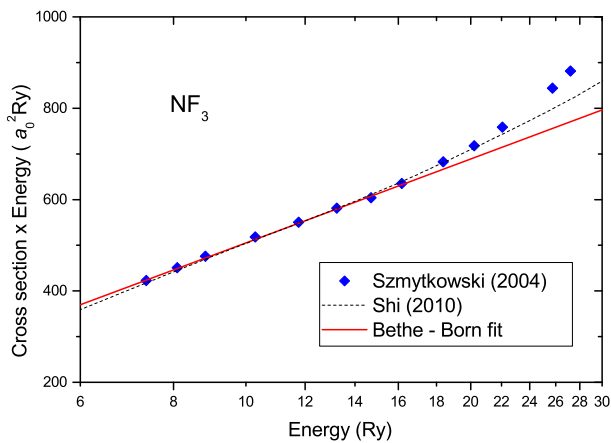}                                
\caption{\label{fig:tot02} Bethe-Born plot of total cross sections by
Szmytkowski {\it et al.} \cite{szmytkowski2004nf3} in their high energy limit.
 Total cross sections from modified additivity rule of Shi {\it et al.} \cite{Shi2010} are also shown for comparison (data read from their fig. \ref{fig:tot01}).}
\end{figure}
Calculations of integral elastic cross section
\cite{hamilton2017nf3, goswami2013cross} predict a resonant structure which is
much narrower (and higher) than the resonance seen in the total cross section
\cite{szmytkowski2004nf3}, see Fig. \ref{fig:tot01}. This may be due to the
neglect of nuclear motion in the calculations. Similar discrepancies between
theory and experiments are observable for molecular targets, like CO$_2$,
N$_2$O: in these molecules the vibrational excitation constitutes a significant
part (about 1/3) of the total cross section \cite{karwasz06}. Calculations for N$_2$O \cite{bettega2006},
 similar to those for 
NF$_3$, also give resonant maxima higher
that the experiment. Note also that NF$_3$ is a polar molecule, so the
interaction with the incoming electron is more attractive in comparison to
targets like CH$_4$, and this shifts maxima to lower energies.
	Two recent calculations \cite{hamilton2017nf3,goswami2013cross} indicate
that the total cross section should rise in the limit of zero energy, due to the
polar character of the molecule. Unfortunately, this was not observed in the
experiment \cite{szmytkowski2004nf3}, probably because the measurements were
stopped at energies higher than the range of such a rise. 
The rather poor angular resolution of Szmytkowski {\it et al.}’s apparatus make
their measurement vulnerable to the angular resolution error at high energies.
To verify this, in Fig. \ref{fig:tot02} we show a Bethe-Born plot of total cross
sections, as done in our previous review in CH$_4$ \cite{song2015cross},
\begin{equation}
\sigma(E) = A/E + B \log (E)/E\,,
\end{equation}
where energy is expressed in Rydbergs, Ry =13.6 eV and the cross sections is
expressed in atomic units $a_0^2 =0.28 \times 10^{-16}$cm$^2$. 
Parameters of the fit, based on experimental points \cite{szmytkowski2004nf3}
between 100-220 eV are $A= -110 \pm 10$ and $B = 610 \pm 20 $.
Contrary to expectations, the plot in Fig.~\ref{fig:tot02} suggests that total
cross sections given by Szmytkowski {\it et al.} \cite{szmytkowski2004nf3} are
overestimates in their high energy limit. We note however that Bethe-Born
analysis is not fully justified at energies of few hundreds of eV, see
discussion in Ref. \cite{Zecca91, Zecca00}. Therefore, in Fig. \ref{fig:tot02}
we plot also total cross sections obtained by the additivity rule
\cite{Shi2010}: these data coincide with the present Bethe-Born fit up to 200 eV and
than deviate slightly upwards. Unfortunately, no information on uncertainties
was given by Shi {\it et al.}\cite{Shi2010}. Table \ref{tab:total} gives our
recommended total cross sections which are based on the experiment of Ref.
\cite{szmytkowski2004nf3} at energies 1-90 eV and on the Bethe-Born fit at
higher energies. 

\begin{table}[!tbp]
\caption{\label{tab:total} Recommended total cross sections (TCS), in $10^{-16}$ cm$^2$ units. In the region 1-100 eV recommended values are adopted from the experiment by Szmytkowski {\it et al.} \cite{szmytkowski2004nf3}. Values at 100-500 eV are obtained from parameters of the Bethe-Born plot, Fig.~\ref{fig:tot02}. The overall uncertainty of  TCS is $\pm$ 10\% at 1-100 eV and 15\% above 100 eV. Additionally from the $\pm$ 10\% uncertainty, TCSs below 1 eV \cite{szmytkowski2004nf3} may be underestimated due to an angular resolution error, by the amount rising with lowering energy. The energy determination is $\pm $0.1eV.}
\begin{tabular}{cccccc}
\hline\hline
Electron& TCS & Electron & TCS& Electron & TCS\\
energy&  & energy&   & energy&  \\
\hline
\hline
0.5 	&	15.9 	&	3.7 	&	25.8 	&	35	&	19.4 	\\
0.6 	&	16.6 	&	4.0 	&	24.9 	&	40	&	19.4 	\\
0.8 	&	18.1 	&	4.5 	&	22.7 	&	45	&	19.4 	\\
1.0 	&	19.6 	&	5.0 	&	20.9 	&	50	&	19.3 	\\
1.2 	&	21.2 	&	5.5 	&	19.5 	&	60	&	18.8 	\\
1.4 	&	22.7 	&	6.0 	&	18.5 	&	70	&	18.3 	\\
1.5 	&	23.3 	&	6.5 	&	17.6 	&	80	&	17.3 	\\
1.6 	&	24.0 	&	7.0 	&	17.2 	&	90	&	16.6 	\\
1.7 	&	24.8 	&	7.5 	&	17.0 	&	100	&	15.9 	\\
1.8 	&	25.2 	&	8.0 	&	16.8 	&	110	&	15.4 	\\
1.9 	&	25.4 	&	8.5 	&	16.7 	&	120	&	14.8 	\\
2.0 	&	26.3 	&	9.0 	&	16.7 	&	140	&	13.8 	\\
2.1 	&	26.9 	&	9.5	&	16.7 	&	160	&	12.9 	\\
2.2 	&	27.4 	&	10	&	16.7 	&	180	&	12.1 	\\
2.3 	&	27.5 	&	11	&	16.6 	&	200	&	11.5 	\\
2.4 	&	27.5 	&	12	&	16.7 	&	220	&	10.9 	\\
2.5 	&	27.7 	&	14	&	16.9 	&	250	&	10.1 	\\
2.6 	&	27.8 	&	16	&	17.2 	&	275	&	9.51 	\\
2.7 	&	28.0 	&	18	&	17.6 	&	300	&	9.00 	\\
2.8 	&	27.9 	&	20	&	17.9 	&	350	&	8.16 	\\
2.9 	&	27.8 	&	22	&	18.2 	&	400	&	7.48 	\\
3.0 	&	27.7 	&	25	&	18.6 	&	450	&	6.91 	\\
3.2 	&	27.2 	&	27	&	18.9 	&	500	&	6.43 	\\
3.5 	&	26.5 	&	30	&	19.2 	&		&		\\
\hline\hline
\end{tabular}
\end{table}

\section{\label{sec:es} Elastic Scattering Cross Section}
\begin{figure}[!tbp]
\includegraphics [width=4cm]{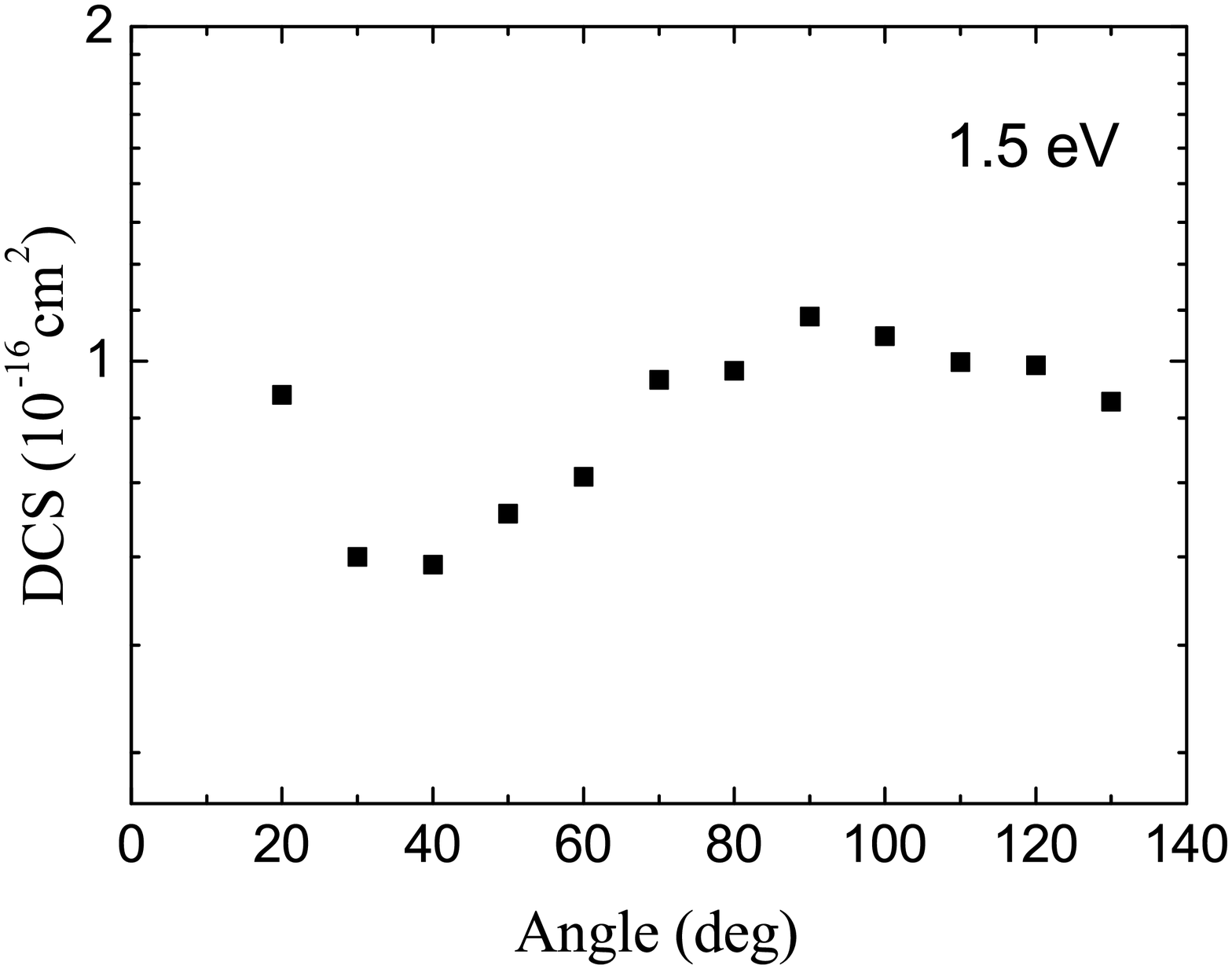}  
\includegraphics [width=4cm]{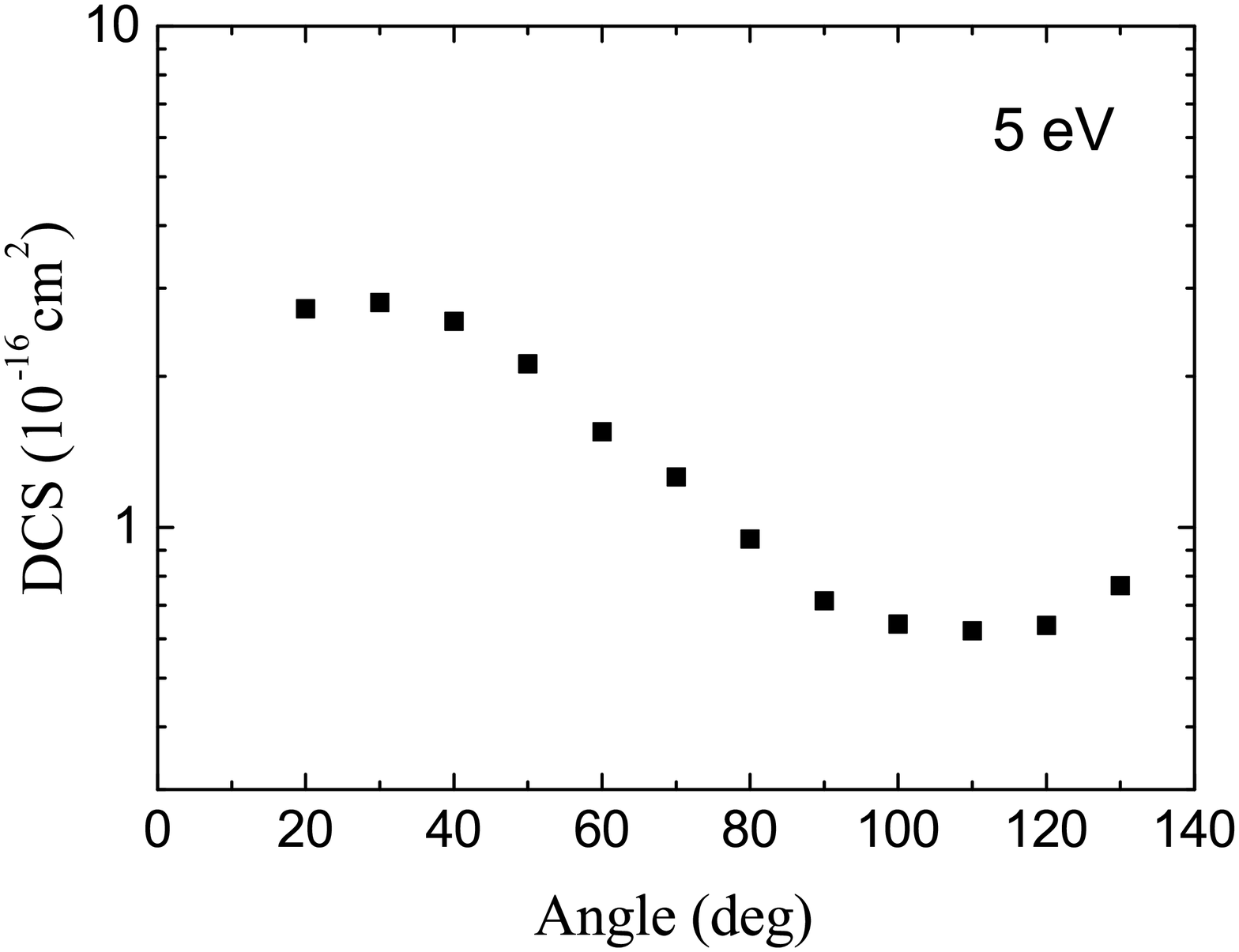}   
\includegraphics [width=4cm]{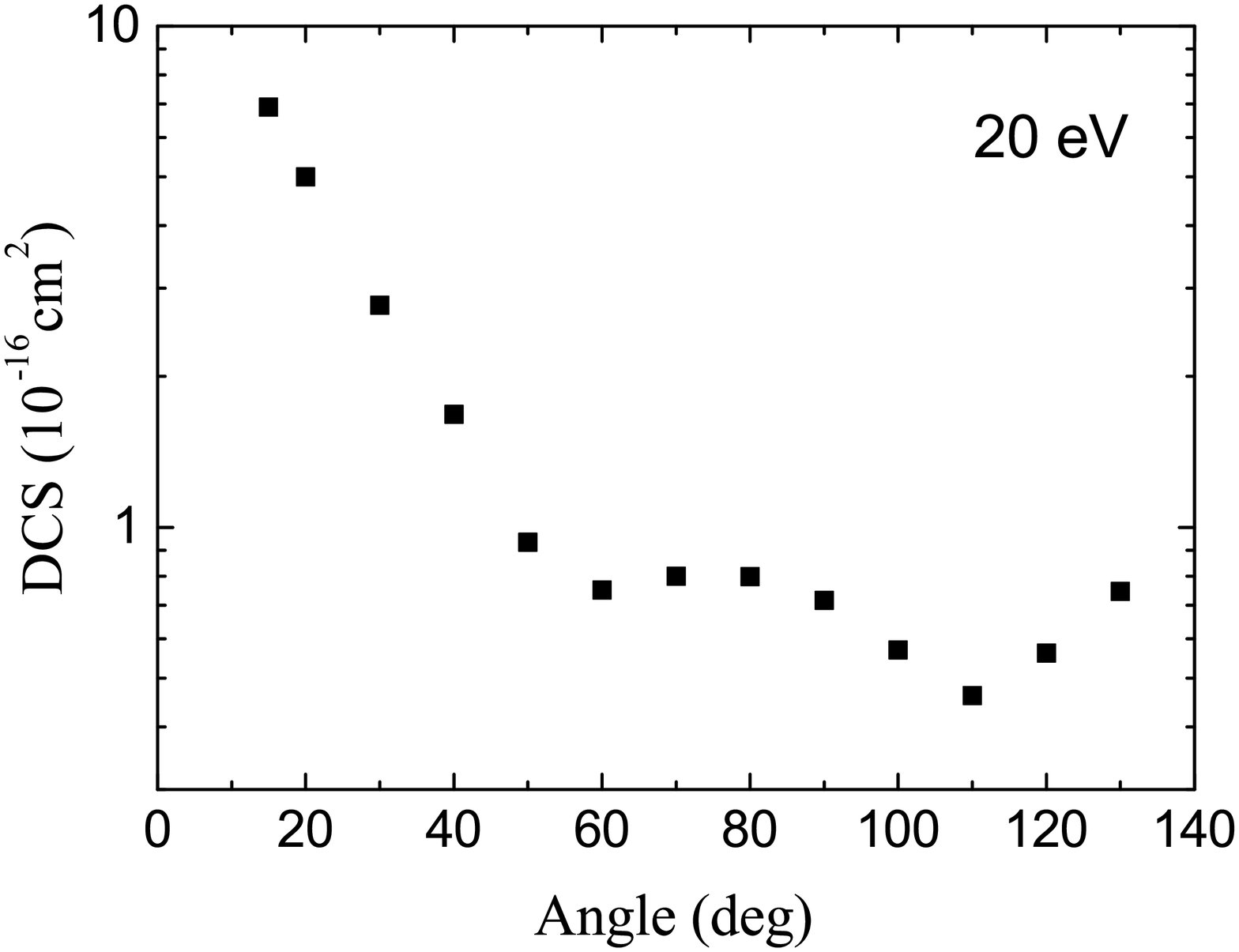}  
\includegraphics [width=4cm]{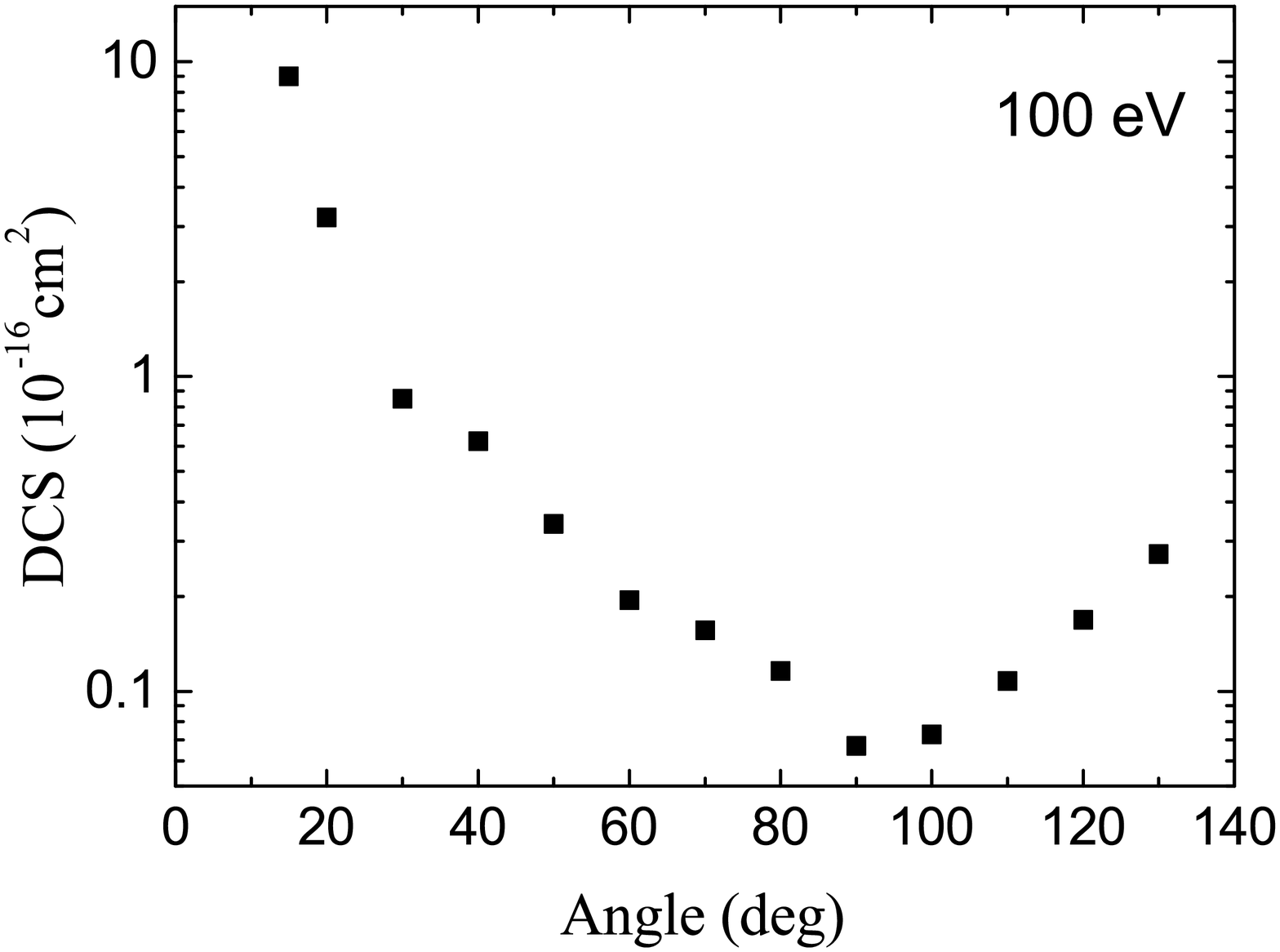}                                        
\caption{\label{fig:es01} Recommended elastic DCS for four representative energies\cite{boesten1996vibrationally}.}
\end{figure}

\begin{figure}[!hbp]
\includegraphics [width=8cm]{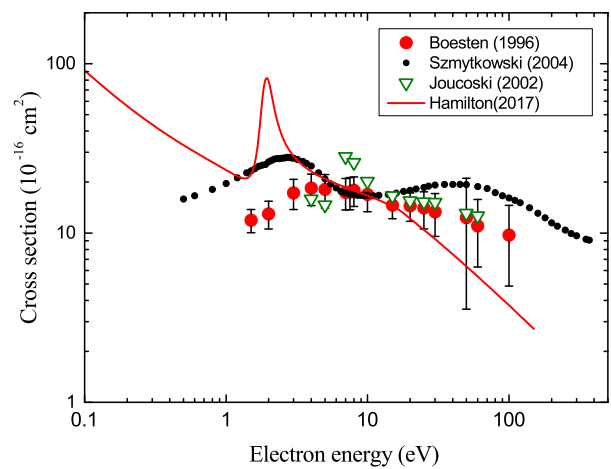}                                
\caption{\label{fig:es02} Recommended elastic integral cross sections with the selected sets of data from the publications\cite{szmytkowski2004nf3,boesten1996vibrationally,joucoski2002elastic,hamilton2017nf3}.}
\end{figure}

Available data for elastic electron scattering from NF$_3$ are very sparse. The
first theoretical study on low-energy electron collision processes in NF$_3$ 
was reported by Rescigno \cite{rescigno1995low} which included Kohn variation
calculations of elastic differential cross sections (DCS) and integral cross
sections (ICS) for electrons with energies in the range 0–10 eV. 
The only comprehensive experimental study, which reported elastic DCS, ICS, and
momentum transfer cross sections (MTCS) for energies between 1.5 and 100 eV and
for angles between 15$^{\circ}$(20$^{\circ}$for energies below 8 eV) and
130$^{\circ}$, was published by Boesten {\it et al.}
\cite{boesten1996vibrationally} Subsequently, a Schwinger multichannel
theoretical approach \cite{joucoski2002elastic} reported corresponding cross
sections for electron energies in the range 0–60 eV. 
Complete numerical values of Boesten {\it et al.}\cite{boesten1996vibrationally}
and 4 representative figures for DCS’s are presented in Table \ref{tab:dcs} and
Fig. \ref{fig:es01}. Their ICS’s are also given in Table \ref{tab:dcs} and Fig.
\ref{fig:es02}. Theoretical ICS’s of Joucoski and Bettega
\cite{joucoski2002elastic} and the total cross sections of Szmytkowski {\it et al.} \cite{szmytkowski2004nf3}  are plotted on the Fig. \ref{fig:es02} for
comparison. 
Generally a good agreement is found between the results from the calculation by
Joucoski and Bettega and the experiment by Boesten {\it et al.}, except a few
points between 5 eV and 10 eV, where the theory exceeds the total cross sections
of Szmytkowski {\it et al.}
The permanent dipole moments of NF$_3$ is 0.0944 au, which is small compared to other polar molecules
such as NH$_3$ (0.578 au) and H$_2$O (0.728 au). Therefore, dipole interaction
between the electron and NF$_3$ would not be important in this collision system
\cite{boesten1996vibrationally}. And, dipole-enhanced forward scattering is
restricted to small angles, usually below 10–20$^{\circ}$, and Boesten {\it et al.}
claimed that this was confirmed in the calculations of
Rescigno\cite{rescigno1995low}  whose DCS at 20$^{\circ}$ reflect their own data. 
In the energy and angular ranges of the experiments by Boesten {\it et al.}, there is no evidence that their
results are unreasonably underestimated, even though there could still be a few
possibilities of slight over- or under-estimation which may not be included in
their uncertainty estimation. So our recommended data is consequently taken from
the measurements of Boesten {\it et al.} Similarly, we recommend their ICS’s.
Boesten {\it et al.} estimated the uncertainties of DCS as 15 \% and of ICS as
30-50 \%. They estimated the contributions of the low and high angle
extrapolations separately, and we present these in their original forms at the
bottom of Table \ref{tab:dcs}. The contributions of the low and high angle
extrapolations are indicated as L\% and R\%, respectively.
 Very recently, Hamilton {\it et al.} published calculated ICS’s\cite{hamilton2017nf3} and are presented in Fig.4 for comparison. 

\begin{table*}[!tbp]
\caption{\label{tab:dcs} Recommended elastic electron scattering cross sections from NF$_3$. DCS’s are in the units of 10$^{-16}$ cm$^2$sr$^{-1}$. Recommended elastic integral cross sections are also given at the bottom in the units of 10$^{-16}$cm$^2$[Ref. 15]. The uncertainties of DCS are 15 \% and of ICS are 30-50 \%. }
\scriptsize
\begin{tabular}{cllllllllllllllll}
\hline\hline
Angle & 1.5eV & 2eV & 3eV & 4eV & 5eV & 7eV & 7.5eV & 8eV & 10eV & 15eV & 20eV & 25eV & 30eV & 50eV & 60eV & 100eV \\
(deg) & DCS & DCS & DCS & DCS & DCS & DCS & DCS & DCS  & DCS & DCS & DCS & DCS & DCS & DCS & DCS & DCS \\
\hline
15 & -- & -- & -- & -- & -- & -- & -- & -- & 3.323 & 4.641 & 6.890 & 9.051 & 10.710 & 12.330 & 11.200 & 9.000 \\
20 & 0.933 & 1.430 & 2.199 & 2.960 & 2.729 & 2.671 & 2.896 & 2.908 & 3.168 & 3.946 & 5.006 & 6.490 & 6.946 & 6.715 & 5.955 & 3.201 \\
30 & 0.667 & 1.216 & 2.436 & 2.949 & 2.807 & 2.932 & 3.036 & 3.132 & 3.037 & 3.077 & 2.777 & 2.863 & 2.657 & 1.838 & 1.243 & 0.851 \\
40 & 0.656 & 1.078 & 2.331 & 2.822 & 2.577 & 2.868 & 2.731 & 2.699 & 2.680 & 2.107 & 1.680 & 1.358 & 1.004 & 0.666 & 0.671 & 0.623 \\
50 & 0.729 & 1.197 & 2.052 & 2.511 & 2.119 & 2.123 & 2.224 & 2.115 & 1.934 & 1.271 & 0.934 & 0.742 & 0.616 & 0.621 & 0.537 & 0.340 \\
60 & 0.787 & 1.152 & 1.768 & 1.818 & 1.552 & 1.655 & 1.517 & 1.460 & 1.390 & 0.826 & 0.750 & 0.688 & 0.639 & 0.601 & 0.328 & 0.195 \\
70 & 0.962 & 1.074 & 1.329 & 1.297 & 1.261 & 1.137 & 1.108 & 1.209 & 0.954 & 0.715 & 0.799 & 0.747 & 0.671 & 0.340 & 0.232 & 0.156 \\
80 & 0.981 & 1.100 & 1.114 & 1.099 & 0.947 & 0.808 & 0.851 & 0.868 & 0.737 & 0.725 & 0.798 & 0.665 & 0.509 & 0.196 & 0.155 & 0.116 \\
90 & 1.097 & 1.011 & 0.920 & 0.794 & 0.714 & 0.727 & 0.719 & 0.766 & 0.702 & 0.786 & 0.715 & 0.510 & 0.320 & 0.116 & 0.109 & 0.067 \\
100 & 1.053 & 0.884 & 0.685 & 0.640 & 0.641 & 0.663 & 0.704 & 0.702 & 0.694 & 0.738 & 0.569 & 0.322 & 0.191 & 0.093 & 0.093 & 0.073 \\
110 & 0.998 & 0.843 & 0.598 & 0.542 & 0.622 & 0.666 & 0.704 & 0.707 & 0.673 & 0.610 & 0.462 & 0.295 & 0.200 & 0.146 & 0.152 & 0.108 \\
120 & 0.992 & 0.778 & 0.584 & 0.539 & 0.637 & 0.652 & 0.661 & 0.639& 0.626 & 0.555 & 0.561 & 0.440 & 0.376 & 0.314 & 0.265 & 0.169 \\
130 & 0.920 & 0.723 & 0.576 & 0.604 & 0.765 & 0.655 & 0.645 & 0.623 & 0.605 & 0.598 & 0.746 & 0.725 & 0.623 & 0.483 & 0.378 & 0.273 \\
ICS & 11.90 & 12.98 & 17.24 & 18.41 & 18.11 & 17.35 & 17.47 & 17.89 & 16.91 & 14.60 & 14.48 & 14.05 & 13.33 & 12.32 & 11.03 & 9.72 \\
L\% & 4 & 5 & 7 & 7 & 8 & 7 & 8 & 13& 6 & 9 & 13 & 18 & 24 & 35 & 41 & 47 \\
R\% & 15 & 18 & 19 & 20 & 21 & 20 & 20 & 15 & 20 & 14 & 14 & 17 & 15 & 62 & 13 & 17 \\

\hline\hline
\end{tabular}
\end{table*}

\section{\label{sec:mt}Momentum Transfer Cross Section}
\begin{figure}[!tbp]
\includegraphics [width=8cm]{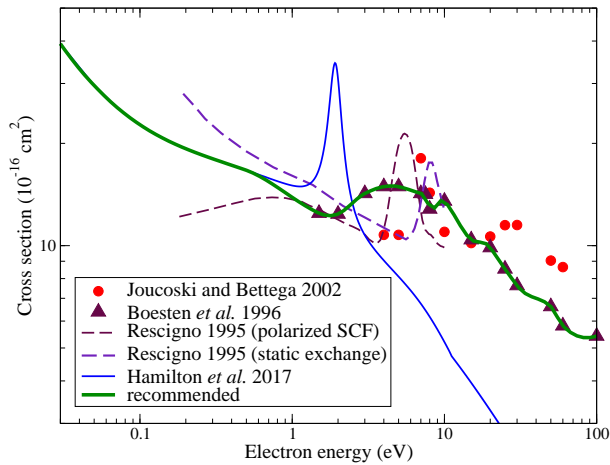}                                
\caption{\label{fig:MTCS}  The momentum-transfer cross section for elastic collisions obtained in different studies. The recommended data are shown by the thick green line. The data is the theoretical results by Hamilton {\it et al.} \cite{hamilton2017nf3}  below 0.65 eV and  the experimental results by Boesten {\it et al.} \cite{boesten1996vibrationally} above 1.5 eV. }
\end{figure}

The momentum-transfer cross section for electron-NF$_3$ collisions was determined in the same studies, mentioned above\cite{boesten1996vibrationally,rescigno1995low,joucoski2002elastic,
hamilton2017nf3}, where the elastic cross sections were measured or computed.
The experimental data by Boesten {\it et al.} \cite{boesten1996vibrationally} is
not complete, especially, at energies below 1~eV. Out of the three theoretical
studies  \cite{rescigno1995low,joucoski2002elastic,hamilton2017nf3}, the most
recent one by Hamilton {\it et al.} \cite{hamilton2017nf3} appears to be the
most accurate one due to a more accurate method (complete active
space-configuration interaction) and a larger basis set employed. However, the
position of the resonance near 1 eV in this study is shifted towards lower
energies compared to the experimental data. The width of the resonance in all
theoretical studies is significantly narrower than in the experiment. Therefore,
at energies above 1~eV, where experimental data exist, we recommend the
experimental data, namely, the one by Boesten {\it et al.}\cite{boesten1996vibrationally} and energies below 1 eV, the theoretical results by Hamilton {\it et al.} \cite{hamilton2017nf3}. The available theoretical and experimental data as well as the recommended set, are shown in Fig. \ref{fig:MTCS}. The values of the recommended data given
in Table \ref{tab:MTCS}. Lisovskiy {\it et al.}\cite{lisovskiy2014electron} measured the 
drift velocity of electrons in NF$_3$ in a limited range of high reduced electric field $E/p$ and 
compared their measurements with the results of the BOLSIG+ calculations using the 
momentum transfer cross sections of Boesten {\it et al.}\cite{boesten1996vibrationally}, 
and Joucoski and Bettega \cite{joucoski2002elastic}. Agreement was satisfactorily, 
especially with calculation using the latter momentum transfer cross section.

\begin{table}
\caption{\label{tab:MTCS} The recommended momentum-transfer cross-section. The data below 0.65 eV are from Hamilton {\it et al.} \cite{hamilton2017nf3} and above 1.5 eV is from Boesten {\it et al.} \cite{boesten1996vibrationally}. Energies are in eV, the cross sections are in units of 10$^{-16}$ cm$^{-1}$.}
\begin{tabular}{llllll}
\hline\hline
Electron& MTCS 	&	Electron	& MTCS 	&Electron 	& MTCS 	\\
Energy&  	&	Energy 	&  	&Energy 	& 	\\
\hline
6.74E-3	&	128.5	&	0.188	&	19.50	&	4.78	&	14.97	\\
7.43E-3	&	117.6	&	0.208	&	19.15	&	5.27	&	14.83	\\
8.20E-3	&	107.7	&	0.229	&	18.82	&	5.82	&	14.63	\\
9.04E-3	&	98.80	&	0.253	&	18.52	&	6.41	&	14.42	\\
9.97E-3	&	90.72	&	0.279	&	18.23	&	7.07	&	14.22	\\
0.0110	&	83.41	&	0.308	&	17.95	&	7.80	&	13.30	\\
0.0121	&	76.80	&	0.339	&	17.68	&	8.61	&	12.91	\\
0.0134	&	70.83	&	0.374	&	17.42	&	9.49	&	13.57	\\
0.0148	&	65.43	&	0.413	&	17.16	&	10.47	&	13.20	\\
0.0163	&	60.56	&	0.455	&	16.91	&	11.55	&	12.34	\\
0.0179	&	56.16	&	0.502	&	16.65	&	12.74	&	11.45	\\
0.0198	&	52.19	&	0.554	&	16.35	&	14.05	&	10.73	\\
0.0218	&	48.61	&	0.611	&	15.99	&	15.49	&	10.32	\\
0.0241	&	45.38	&	0.674	&	15.60	&	17.09	&	10.24	\\
0.0266	&	42.47	&	0.743	&	15.17	&	18.85	&	10.13	\\
0.0293	&	39.85	&	0.820	&	14.74	&	20.79	&	9.63	\\
0.0323	&	37.49	&	0.904	&	14.31	&	22.93	&	9.03	\\
0.0356	&	35.36	&	1.00	&	13.89	&	25.29	&	8.48	\\
0.0393	&	33.46	&	1.10	&	13.49	&	27.89	&	7.98	\\
0.0434	&	31.75	&	1.21	&	13.12	&	30.76	&	7.52	\\
0.0478	&	30.23	&	1.34	&	12.79	&	33.93	&	7.24	\\
0.0527	&	28.88	&	1.48	&	12.50	&	37.42	&	7.08	\\
0.0582	&	27.64	&	1.63	&	12.30	&	41.27	&	6.97	\\
0.0642	&	26.50	&	1.79	&	12.25	&	45.52	&	6.83	\\
0.0708	&	25.50	&	1.98	&	12.37	&	50.21	&	6.61	\\
0.0780	&	24.60	&	2.18	&	12.69	&	55.38	&	6.15	\\
0.0861	&	23.78	&	2.41	&	13.17	&	61.08	&	5.76	\\
0.0949	&	23.05	&	2.66	&	13.70	&	67.36	&	5.56	\\
0.105	&	22.38	&	2.93	&	14.16	&	74.30	&	5.43	\\
0.115	&	21.78	&	3.23	&	14.46	&	81.95	&	5.37	\\
0.127	&	21.24	&	3.56	&	14.72	&	90.38	&	5.36	\\
0.140	&	20.74	&	3.93	&	14.90	&	99.69	&	5.42	\\
0.155	&	20.29	&	4.33	&	14.98	&	109.95	&	5.50	\\
0.171	&	19.88	&		&		&		&		\\
\hline\hline
\end{tabular}
\end{table}

\section{\label{sec:ro}Rotational Excitation Cross Section}
Due to its $C_{3v}$ symmetry at equilibrium geometry, NF$_3$ is a 
symmetric top in the rigid-rotor approximation. It is an oblate 
rotor with rotational constants given in Table \ref{tab:properties}. As for 
other symmetric top molecules, the rotational levels of NF$_3$ are characterized 
by two quantum numbers, the rotational angular momentum $j$ and its projection 
$k$ on the molecular symmetry axis.  The fluorine atom has only one stable 
isotope, $^{19}$F with nuclear spin $i=1/2$. Therefore, the total nuclear spin 
of three fluorine atoms could be $I=1/2$ (para-NF$_3$) or $3/2$ (ortho-NF$_3$). 
In the following discussion, we neglect the hyper-fine interaction and mixing 
between singlet and triplet nuclear-spin states of NF$_3$. The total wave 
function, including the nuclear-spin part, of NF$_3$ should be of the $A_2$ 
irreducible representation of the $C_{3v}$ group because $^{19}$F is a fermion. 
It means that for ortho-NF$_3$, the space part (rovibronic) of the wave function 
should also be of  the $A_2$ irreducible representation, because the nuclear 
spin part is totally symmetric, $A_1$. For para-NF$_3$, the space part of the 
wave function should be of the $E$ irreducible representation.
In the both cases, it leads to the conclusion that the lowest allowed rotational level in the 
ground {\it vibronic} state has $j=1$. The $j=0$ rotational level is forbidden 
for  the ground {\it vibronic} state because the  $j=0$ rotational level is of 
the $A_1$ representation. For certain excited vibrational or/and electronic 
states of $E$ and $A_2$ representations of the $\nu_3$ and $\nu_4$ modes, the 
$j=0$ rotational level is allowed.

\begin{figure}[!tbp]                               
\includegraphics [width=8cm]{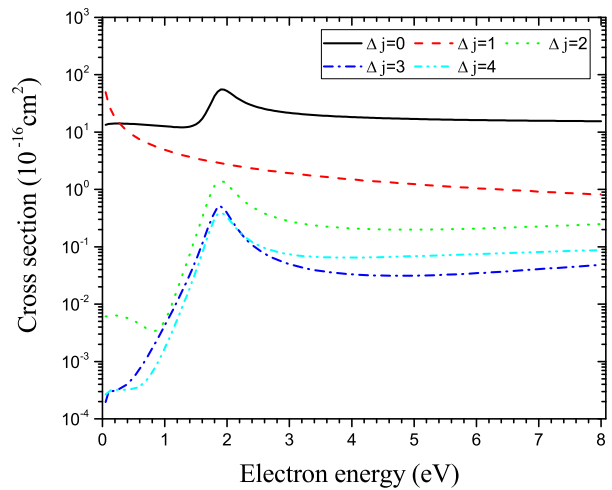}                                  
\caption{\label{fig:rota_exc} Rotational excitation cross section from the ground rotational level $j=1$ to the $j'=1-5$ levels.}
\end{figure}

The only published data on rotational excitation is a theoretical
calculation by Goswami {\it et al.} \cite{goswami2013cross}, where
rotational excitation cross sections starting from $j=0$ were
calculated using the UK R-matrix code and the Quantemol interface
\cite{tennyson07quantemol}. In order to account for transitions starting from a $j=1$ rotational ground state,
we employed a similar procedure using the scattering wave functions of
Hamilton {\it et al.} \cite{hamilton2017nf3} and, in the outer region,
the experimental value for the NF$_3$ dipole moment.
Our new data are reproduced in
Fig.~\ref{fig:rota_exc} and numerical
values are given in Table \ref{tab:rot_ex}. 
The magnitudes of the $\Delta j = 0$ transition cross sections
presented here are similar those calculated by Goswami {\it et al.}
\cite{goswami2013cross}. The main differences arise in region of the
two shape resonances.  The location of the resonances features in
Fig.~\ref{fig:rota_exc} are in agreement with a measured values of
Nandi {\it et al.}  \cite{nandi2001nf3} at 1.855 -- 1.914 eV; the
resonances of Goswami {\it et al.} are placed higher, at around 4 eV.  The
dipole-allowed, $\Delta$ $j = 1$ cross sections calculated in this
work are larger than those of Goswami {\it et al.} This is due the
different initial $j$, and subsequent $j'$, states considered.  Our
$\Delta J > 1$ transition cross sections are of similar magnitude to
those calculated by Goswami {\it et al.}, except when they are
affected by the location of the shape resonances.

We also performed a
quick calculation to estimate the uncertainty of the obtained cross
sections due to parameters of the quantum-chemistry model used.
The estimated uncertainty is about 5\%
for elastic $\Delta j=0$ transition, and about 20\% for inelastic
transitions.

\begin{table}[!tbp]
\caption{\label{tab:rot_ex} 
The recommended cross sections for rotational excitation from the ground rotational level $j=1$ to the $j'=1-5$ levels. Energies are in eV, the cross sections are in units of 10$^{-16}$ cm$^{-1}$.}
\begin{tabular}{lllllll}
\hline\hline
Electron &  CS	&CS 	& CS 	&CS	& CS 	\\
Energ & $\Delta j=0$ &$\Delta j=1$	& $\Delta j=2$ 	&  $\Delta j=3$ 	&  $\Delta j=4$ \\
\hline
0.05 	&	13.41 	&	49.59 	&	6.115E-03	&	1.960E-04	&	2.643E-04	\\
0.10 	&	13.85 	&	30.61 	&	6.176E-03	&	3.019E-04	&	3.130E-04	\\
0.15 	&	14.01 	&	22.61 	&	6.293E-03	&	3.038E-04	&	3.169E-04	\\
0.20 	&	14.07 	&	18.11 	&	6.325E-03	&	3.113E-04	&	3.192E-04	\\
0.25 	&	14.08 	&	15.20 	&	6.275E-03	&	3.239E-04	&	3.202E-04	\\
0.30 	&	14.06 	&	13.15 	&	6.155E-03	&	3.428E-04	&	3.206E-04	\\
0.35 	&	14.02 	&	11.62 	&	5.974E-03	&	3.700E-04	&	3.213E-04	\\
0.40 	&	13.97 	&	10.44 	&	5.740E-03	&	4.080E-04	&	3.231E-04	\\
0.45 	&	13.90 	&	9.48 	&	5.462E-03	&	4.599E-04	&	3.274E-04	\\
0.50 	&	13.82 	&	8.70 	&	5.149E-03	&	5.298E-04	&	3.357E-04	\\
0.55 	&	13.74 	&	8.05 	&	4.813E-03	&	6.224E-04	&	3.502E-04	\\
0.60 	&	13.64 	&	7.49 	&	4.467E-03	&	7.439E-04	&	3.735E-04	\\
0.65 	&	13.53 	&	7.01 	&	4.128E-03	&	9.016E-04	&	4.094E-04	\\
0.70 	&	13.42 	&	6.60 	&	3.820E-03	&	1.105E-03	&	4.625E-04	\\
0.75 	&	13.30 	&	6.23 	&	3.572E-03	&	1.366E-03	&	5.394E-04	\\
0.80 	&	13.18 	&	5.90 	&	3.422E-03	&	1.699E-03	&	6.484E-04	\\
0.85 	&	13.05 	&	5.61 	&	3.423E-03	&	2.125E-03	&	8.009E-04	\\
0.90 	&	12.92 	&	5.35 	&	3.642E-03	&	2.667E-03	&	1.012E-03	\\
0.95 	&	12.78 	&	5.11 	&	4.173E-03	&	3.358E-03	&	1.303E-03	\\
1.00 	&	12.64 	&	4.90 	&	5.142E-03	&	4.242E-03	&	1.701E-03	\\
1.05 	&	12.51 	&	4.70 	&	6.719E-03	&	5.374E-03	&	2.244E-03	\\
1.10 	&	12.38 	&	4.52 	&	9.142E-03	&	6.832E-03	&	2.986E-03	\\
1.15 	&	12.26 	&	4.35 	&	1.274E-02	&	8.718E-03	&	4.002E-03	\\
1.20 	&	12.16 	&	4.20 	&	1.799E-02	&	1.118E-02	&	5.395E-03	\\
1.25 	&	12.10 	&	4.06 	&	2.554E-02	&	1.441E-02	&	7.317E-03	\\
1.30 	&	12.10 	&	3.93 	&	3.637E-02	&	1.871E-02	&	9.987E-03	\\
1.40 	&	12.40 	&	3.69 	&	7.420E-02	&	3.237E-02	&	1.902E-02	\\
1.50 	&	13.63 	&	3.48 	&	1.533E-01	&	5.857E-02	&	3.763E-02	\\
1.60 	&	17.12 	&	3.29 	&	3.218E-01	&	1.120E-01	&	7.771E-02	\\
1.70 	&	25.92 	&	3.13 	&	6.627E-01	&	2.232E-01	&	1.637E-01	\\
1.80 	&	42.51 	&	2.99 	&	1.160E+00	&	4.090E-01	&	3.081E-01	\\
1.85 	&	50.83 	&	2.93 	&	1.343E+00	&	4.859E-01	&	3.701E-01	\\
1.90 	&	55.37 	&	2.87 	&	1.404E+00	&	5.021E-01	&	3.911E-01	\\
1.95 	&	55.23 	&	2.80 	&	1.349E+00	&	4.579E-01	&	3.697E-01	\\
2.00 	&	52.04 	&	2.74 	&	1.226E+00	&	3.873E-01	&	3.266E-01	\\
2.05 	&	47.80 	&	2.68 	&	1.083E+00	&	3.185E-01	&	2.808E-01	\\
2.10 	&	43.65 	&	2.62 	&	9.487E-01	&	2.617E-01	&	2.408E-01	\\
2.20 	&	37.01 	&	2.51 	&	7.369E-01	&	1.836E-01	&	1.824E-01	\\
2.30 	&	32.45 	&	2.42 	&	5.942E-01	&	1.372E-01	&	1.458E-01	\\
2.40 	&	29.30 	&	2.33 	&	4.978E-01	&	1.084E-01	&	1.223E-01	\\
2.50 	&	27.06 	&	2.25 	&	4.306E-01	&	8.935E-02	&	1.065E-01	\\
2.60 	&	25.40 	&	2.17 	&	3.822E-01	&	7.618E-02	&	9.543E-02	\\
2.70 	&	24.13 	&	2.10 	&	3.461E-01	&	6.666E-02	&	8.750E-02	\\
2.80 	&	23.12 	&	2.04 	&	3.185E-01	&	5.955E-02	&	8.166E-02	\\
2.90 	&	22.31 	&	1.98 	&	2.969E-01	&	5.409E-02	&	7.728E-02	\\
3.00 	&	21.64 	&	1.92 	&	2.796E-01	&	4.981E-02	&	7.396E-02	\\
3.50 	&	19.49 	&	1.68 	&	2.302E-01	&	3.789E-02	&	6.611E-02	\\
4.00 	&	18.29 	&	1.50 	&	2.095E-01	&	3.314E-02	&	6.490E-02	\\
4.50 	&	17.50 	&	1.35 	&	2.010E-01	&	3.140E-02	&	6.623E-02	\\
5.00 	&	16.93 	&	1.23 	&	1.993E-01	&	3.136E-02	&	6.863E-02	\\
5.50 	&	16.52 	&	1.13 	&	2.020E-01	&	3.250E-02	&	7.150E-02	\\
6.00 	&	16.20 	&	1.05 	&	2.079E-01	&	3.454E-02	&	7.456E-02	\\
6.50 	&	15.97 	&	0.98 	&	2.161E-01	&	3.731E-02	&	7.774E-02	\\
7.00 	&	15.78 	&	0.92 	&	2.260E-01	&	4.065E-02	&	8.102E-02	\\
7.50 	&	15.62 	&	0.87 	&	2.369E-01	&	4.447E-02	&	8.443E-02	\\
8.00 	&	15.47 	&	0.82 	&	2.484E-01	&	4.867E-02	&	8.799E-02	\\
\hline\hline
\end{tabular}
\end{table}

\section{\label{sec:vi}Vibrational Excitation Cross Sections}
The NF$_3$ molecule has four vibrational  modes, two of which are $A_1$ 
non-degenerate modes, $\nu_1$ and $\nu_2$, and the two others, $\nu_3$ and 
$\nu_4$, are doubly-degenerate of  $E$ symmetry 
\cite{boesten1996vibrationally}. The excitation energies of the modes are given 
in Table \ref{tab:modes}.

\begin{table}[!hbp]
\caption{\label{tab:modes}Vibrational modes and excitation energies of NF$_3$\cite{boesten1996vibrationally}. }
\begin{tabular}{ccc}
\hline\hline
Mode   & Type & Energy (eV) \\
\hline
$\nu_1$	& Symmetric stretch &	0.1280\\
$\nu_2$	& Umbrella mode &	0.0802\\
$\nu_3$	& Asymmetric stretch	& 0.1125\\
$\nu_4$	& Asymmetric bend	& 0.0611\\
\hline\hline
\end{tabular}
\end{table}

\begin{figure}[!tbp]                                         
\includegraphics [width=8cm]{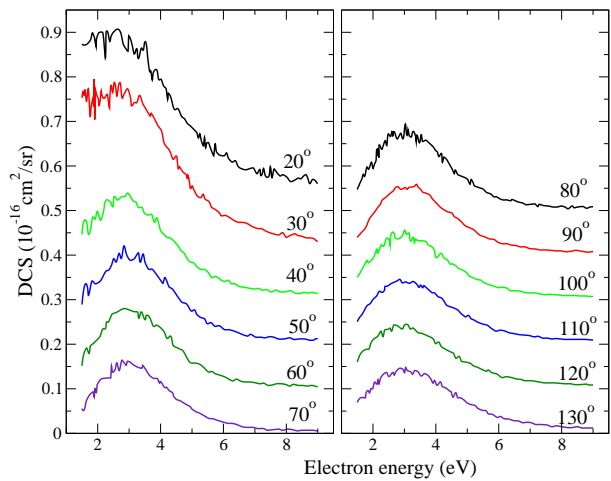}                                  
\caption{\label{fig:dcs} Experimental differential cross section for excitation of the $\nu_1/\nu_3$ modes \cite{boesten1996vibrationally}. The values for $60^\circ$, $70^\circ$, etc. are shifted by 0.1, 0.2, etc. units and approach zero at 9 eV. }
\end{figure}

\begin{figure}                                          
\includegraphics [width=8cm]{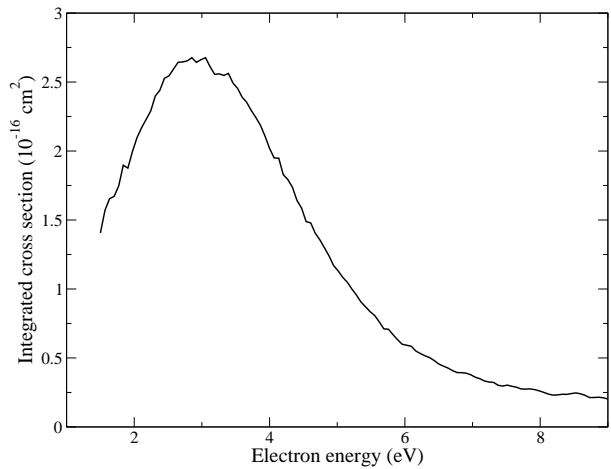}                                  
\caption{\label{fig:vib_ex_integ} Integrated cross section for excitation of the $\nu_1/\nu_3$ modes obtained from the data shown in Fig.~\ref{fig:dcs}\cite{boesten1996vibrationally}. }
\end{figure}

The only available experimental data on vibrational excitation is by Boesten 
{\it et al.}, \cite{boesten1996vibrationally} where differential cross sections 
for excitation of the $\nu_1/\nu_3$ modes were measured in a crossed-beam 
experiment. The cross sections obtained are reproduced in Fig.~\ref{fig:dcs}. 
Figure \ref{fig:vib_ex_integ} shows the cross section integrated over the solid 
angle. The corresponding numerical values are given in Table~\ref{tab:vibr_exc}. 
Because the differential cross section was not measured for angles below 
$20^\circ$ and above $130^\circ$, performing the integration, we assumed that 
the DCS below $20^\circ$ is equal to the one at $20^\circ$ and the DCS above 
$130^\circ$ is equal to the one at $130^\circ$. Such an assumption introduces a 
significant uncertainty, of the order of 30\%, into the integrated cross 
section. Note that the present estimate of the vibrational cross section at 2.5 
eV maximum agrees with the value used by Lisovskiy {\it et al.} 
\cite{lisovskiy2014electron} for modeling electron transport coefficients in 
NF$_3$. 


\begin{table}
\caption{\label{tab:vibr_exc} Integrated cross section for vibrational excitation of the $\nu_1/\nu_3$ modes \cite{boesten1996vibrationally}. Electron energies are in eV, the cross sections are in units of 10$^{-16}$ cm$^{-1}$.}
\begin{tabular}{cccccc}
\hline\hline
Electron& CS&	Electron 	& CS	&Electron  	& CS 	\\
Energy& &	Energy 	& 	&Energy 	&  	\\
\hline
1.50	&	1.41	&	4.07	&	1.95	&	6.57	&	0.44	\\
1.57	&	1.57	&	4.14	&	1.95	&	6.64	&	0.43	\\
1.64	&	1.65	&	4.20	&	1.83	&	6.70	&	0.41	\\
1.70	&	1.67	&	4.27	&	1.79	&	6.77	&	0.39	\\
1.77	&	1.75	&	4.34	&	1.74	&	6.84	&	0.39	\\
1.84	&	1.90	&	4.41	&	1.64	&	6.91	&	0.39	\\
1.91	&	1.88	&	4.47	&	1.59	&	6.97	&	0.38	\\
1.97	&	1.99	&	4.54	&	1.49	&	7.04	&	0.36	\\
2.04	&	2.10	&	4.61	&	1.48	&	7.11	&	0.35	\\
2.11	&	2.17	&	4.68	&	1.40	&	7.18	&	0.33	\\
2.18	&	2.23	&	4.74	&	1.36	&	7.24	&	0.33	\\
2.24	&	2.29	&	4.81	&	1.30	&	7.31	&	0.32	\\
2.31	&	2.40	&	4.88	&	1.24	&	7.38	&	0.30	\\
2.38	&	2.44	&	4.95	&	1.17	&	7.45	&	0.30	\\
2.45	&	2.53	&	5.01	&	1.13	&	7.51	&	0.30	\\
2.51	&	2.54	&	5.08	&	1.08	&	7.58	&	0.29	\\
2.58	&	2.59	&	5.15	&	1.05	&	7.65	&	0.29	\\
2.65	&	2.64	&	5.22	&	1.00	&	7.72	&	0.28	\\
2.72	&	2.65	&	5.28	&	0.96	&	7.78	&	0.27	\\
2.78	&	2.65	&	5.35	&	0.91	&	7.85	&	0.28	\\
2.85	&	2.68	&	5.42	&	0.87	&	7.92	&	0.27	\\
2.92	&	2.64	&	5.49	&	0.84	&	7.99	&	0.26	\\
2.99	&	2.66	&	5.55	&	0.81	&	8.05	&	0.25	\\
3.05	&	2.68	&	5.62	&	0.76	&	8.12	&	0.24	\\
3.12	&	2.61	&	5.69	&	0.71	&	8.19	&	0.23	\\
3.19	&	2.56	&	5.76	&	0.71	&	8.26	&	0.23	\\
3.26	&	2.56	&	5.82	&	0.67	&	8.32	&	0.24	\\
3.32	&	2.55	&	5.89	&	0.63	&	8.39	&	0.24	\\
3.39	&	2.56	&	5.96	&	0.60	&	8.46	&	0.24	\\
3.46	&	2.49	&	6.03	&	0.59	&	8.53	&	0.25	\\
3.53	&	2.45	&	6.09	&	0.58	&	8.59	&	0.24	\\
3.59	&	2.39	&	6.16	&	0.55	&	8.66	&	0.23	\\
3.66	&	2.35	&	6.23	&	0.53	&	8.73	&	0.21	\\
3.73	&	2.29	&	6.30	&	0.52	&	8.80	&	0.21	\\
3.80	&	2.24	&	6.36	&	0.50	&	8.86	&	0.22	\\
3.86	&	2.19	&	6.43	&	0.48	&	8.93	&	0.21	\\
3.93	&	2.11	&	6.50	&	0.46	&	9.00	&	0.20	\\
4.00	&	2.02	&		&		&		&		\\

\hline\hline
\end{tabular}
 \\
\end{table}
\section{Electron impact electronic excitation and dissociation}
There are no experimental determinations of electron impact electronic excitation or
dissociation. Theoretically Goswami {\it et al} \cite{goswami2013cross} considered
inelastic processes in their  spherical complex optical potential (SCOP) calculations but
this procedure does not separate these into their individual contributions. Here we therefore concentrate
on the recent R-matrix calculation \cite{jt518} by Hamilton {\it et al} \cite{hamilton2017nf3} 
and older Kohn calculations by Rescigno \cite{rescigno1995low}. Both calculations are based on
the use of a close-coupling expansion of the target wave functions.

Electron impact dissociation reactions go via excitation to electronically excited states of the target which then dissociate \cite{jt229}.
The dissociation of  the N-F in NF$_3$  is 2.52 eV \cite{vedeneyev1962energiya,kennedy1961strength} and no
low-lying metastable electronically-excited states of NF$_3$ are known. 
It can therefore be assumed that all electronic excitation leads to dissociation;
a similar assumption has been made in cases where the results are testable
against experiment \cite{jt585} and
has been found to be reasonable. Both Rescigno and Hamilton {\it et al} made this assumption for NF$_3$.

For an accurate calculation of these processes a large number of electronically
excited states need to be considered.
Born corrections to the electron-impact excitation cross sections were used to
account for long range dipole effects \cite{chu1974rotational,jt256}.

\begin{figure}[!tbp] 
\includegraphics [width=9cm]{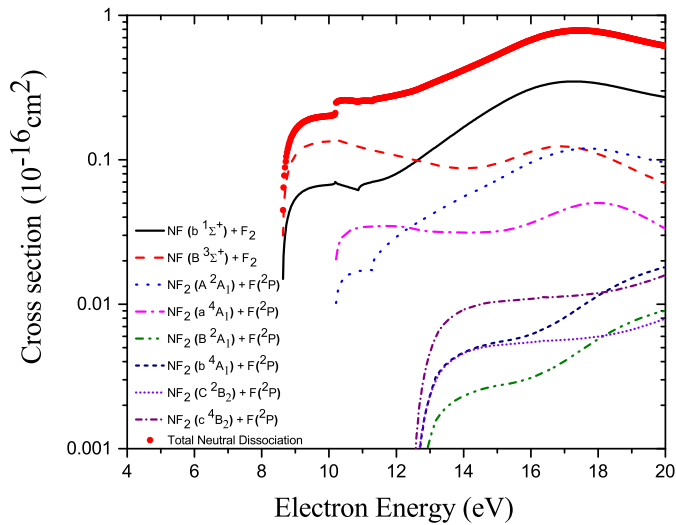}                                
\caption{\label{fig:ex}Cross sections for electron impact dissociation into various channels, taken from 
the recent R-matrix calculation by Hamilton {\it et al} \cite{hamilton2017nf3}}
\end{figure}

Hamilton {\it et al} \cite{hamilton2017nf3} estimated the products of the dissociation
process by analogy with the observed  photodissociation cross sections of Seccombe {\it et al.}  
\cite{Seccombe2001} which suggested that the following process can occur:
\begin{equation}
\mathrm{NF}_{3}(a\enskip^{1}E) \rightarrow \mathrm{NF}(b\enskip^{1}\Sigma^{+}) + \mathrm{F}_{2}
\label{ref:NF3_D_1}
\end{equation}
\begin{equation}
\mathrm{NF}_{3}(b\enskip^{1}A_{1}) \rightarrow \mathrm{NF}_{2}(A\enskip^{2}A_{1}) + \mathrm{F}(^{2}P_{u})
\label{ref:NF3_D_2}
\end{equation}
\begin{equation}
\mathrm{NF}_{3}(c\enskip^{1}A_{2}) \rightarrow \mathrm{NF}_{2}(B\enskip^{2}A_{1}) + \mathrm{F}(^{2}P_{u})
\label{ref:NF3_D_3}
\end{equation}
\begin{equation}
\mathrm{NF}_{3}(d\enskip^{1}E) \rightarrow \mathrm{NF}_{2}(C\enskip^{2}B_{2}) + \mathrm{F}(^{2}P_{u})
\label{ref:NF3_D_4}
\end{equation}
\begin{equation}
\mathrm{NF}_{3}(A\enskip^{3}E) \rightarrow \mathrm{NF}(b\enskip^{1}\Sigma^{+} ) + \mathrm{F}_{2}
\end{equation}
\begin{equation}
\mathrm{NF}_{3}(A\enskip^{3}E) \rightarrow \mathrm{NF}(B\enskip^{3}\Sigma^{+}) + \mathrm{F}_{2}
\label{ref:NF3_D_6}
\end{equation}
\begin{equation}
\mathrm{NF}_{3}(B\enskip^{3}A_{1}) \rightarrow \mathrm{NF}_{2}(A\enskip^{2}A_{1}) + \mathrm{F}(^{2}P_{u})
\end{equation}
\begin{equation}
\mathrm{NF}_{3}(B\enskip^{3}A_{1}) \rightarrow \mathrm{NF}_{2}(a\enskip^{4}A_{1}) + \mathrm{F}(^{2}P_{u})
\label{ref:NF3_D_7}
\end{equation}
\begin{equation}
\mathrm{NF}_{3}(C\enskip^{3}A_{2}) \rightarrow \mathrm{NF}_{2}(B\enskip^{2}A_{1}) + \mathrm{F}(^{2}P_{u}) 
\end{equation}
\begin{equation}
\mathrm{NF}_{3}(C\enskip^{3}A_{2}) \rightarrow \mathrm{NF}_{2}(b\enskip^{4}A_{1}) + \mathrm{F}(^{2}P_{u})
\label{ref:NF3_D_8}
\end{equation}
\begin{equation}
\mathrm{NF}_{3}(D\enskip^{3}E) \rightarrow \mathrm{NF}_{2}(C\enskip^{2}B_{2} ) + \mathrm{F}(^{2}P_{u})
\end{equation}
\begin{equation}
\mathrm{NF}_{3}(D\enskip^{3}E) \rightarrow \mathrm{NF}_{2}(c\enskip^{4}B_{2}) + \mathrm{F}(^{2}P_{u})
\label{ref:NF3_D_9}
\end{equation} 
These results, which are given in Figure~\ref{fig:ex} and Table~\ref{tab:diss}, are the best currently 
available but they must be considered to be estimates. Experimental studies of 
electron impact dissociation of NF$_3$ would
be very useful.

\begin{table*}[!tbp] 
\caption{\label{tab:diss}Cross sections for electron impact dissociation calculated by  Hamilton {\it et al.} \cite{hamilton2017nf3};the products of the dissociation process were estimated  by analogy with the observed  photodissociation cross sections of Seccombe {\it et al.}  \cite{Seccombe2001}}
\begin{tabular}{lccclccc}
\hline
\hline
Energy	&	Neutral Dissociation	&	NF +F$_2$ &	NF$_2$+F & Energy	&	Neutral Dissociation	&	NF+F$_2$ &	NF$_2$+F\\
(eV)	&	($10^{-16} cm^2$)&	($10^{-16} cm^2$)&	($10^{-16} cm^2$)&(eV)	&	($10^{-16} cm^2$)&	($10^{-16} cm^2$)&	($10^{-16} cm^2$)\\
\hline
8.63 	&	0.04487	&	0.04487	&		&	10.91 	&	0.25451	&	0.18656	&	0.05096	\\
8.74 	&	0.11003	&	0.11003	&		&	10.95 	&	0.25555	&	0.18735	&	0.05115	\\
8.86 	&	0.13913	&	0.14	&		&	10.99 	&	0.25605	&	0.18763	&	0.05131	\\
8.97 	&	0.15767	&	0.16	&		&	11.01 	&	0.2562	&	0.18769	&	0.05138	\\
9.09 	&	0.17047	&	0.17	&		&	11.05 	&	0.25636	&	0.18768	&	0.05151	\\
9.20 	&	0.17963	&	0.18	&		&	11.09 	&	0.2564	&	0.18758	&	0.05161	\\
9.32 	&	0.18627	&	0.19	&		&	11.11 	&	0.25638	&	0.18751	&	0.05166	\\
9.43 	&	0.19111	&	0.19	&		&	11.15 	&	0.2563	&	0.18732	&	0.05173	\\
9.54 	&	0.19459	&	0.19	&		&	11.19 	&	0.25616	&	0.18711	&	0.05178	\\
9.66 	&	0.19707	&	0.20	&		&	11.21 	&	0.25607	&	0.18701	&	0.0518	\\
9.77 	&	0.19884	&	0.20	&		&	11.25 	&	0.25587	&	0.1868	&	0.05181	\\
9.89 	&	0.20012	&	0.20	&		&	11.27 	&	0.25575	&	0.1867	&	0.05179	\\
10.00 	&	0.20125	&	0.20	&		&	11.29 	&	0.2556	&	0.18663	&	0.05173	\\
10.01 	&	0.20136	&	0.20136	&		&	11.31 	&	0.25913	&	0.18676	&	0.05332	\\
10.05 	&	0.20186	&	0.20186	&		&	11.35 	&	0.26174	&	0.18636	&	0.05495	\\
10.09 	&	0.20252	&	0.20252	&		&	11.39 	&	0.26303	&	0.18612	&	0.05575	\\
10.11 	&	0.20296	&	0.20296	&		&	11.41 	&	0.26359	&	0.18602	&	0.0561	\\
10.15 	&	0.20439	&	0.20439	&		&	11.45 	&	0.26463	&	0.18582	&	0.05675	\\
10.17 	&	0.20583	&	0.20583	&		&	11.49 	&	0.26562	&	0.18566	&	0.05735	\\
10.19 	&	0.21064	&	0.21064	&		&	11.50 	&	0.26587	&	0.18562	&	0.0575	\\
10.21 	&	0.24834	&	0.20752	&	0.03062	&	11.89 	&	0.27628	&	0.18605	&	0.06249	\\
10.23 	&	0.25086	&	0.20595	&	0.03368	&	12.27 	&	0.2898	&	0.19053	&	0.06671	\\
10.25 	&	0.25271	&	0.20493	&	0.03583	&	12.66 	&	0.30896	&	0.19843	&	0.0735	\\
10.29 	&	0.25521	&	0.20332	&	0.03892	&	13.05 	&	0.33706	&	0.21021	&	0.08497	\\
10.31 	&	0.25608	&	0.2026	&	0.04011	&	13.43 	&	0.36743	&	0.22496	&	0.09535	\\
10.35 	&	0.25731	&	0.20125	&	0.04205	&	13.82 	&	0.40073	&	0.24445	&	0.10366	\\
10.39 	&	0.25805	&	0.19994	&	0.04358	&	14.20 	&	0.43873	&	0.26877	&	0.11144	\\
10.41 	&	0.25828	&	0.1993	&	0.04424	&	14.59 	&	0.48114	&	0.29735	&	0.11912	\\
10.45 	&	0.25853	&	0.19803	&	0.04537	&	14.98 	&	0.52912	&	0.32951	&	0.12768	\\
10.49 	&	0.25854	&	0.19678	&	0.04632	&	15.36 	&	0.58334	&	0.36512	&	0.1378	\\
10.51 	&	0.25847	&	0.19615	&	0.04674	&	15.75 	&	0.64031	&	0.40123	&	0.14951	\\
10.55 	&	0.25822	&	0.19491	&	0.04748	&	16.14 	&	0.69457	&	0.43312	&	0.16253	\\
10.59 	&	0.25784	&	0.19368	&	0.04812	&	16.52 	&	0.73877	&	0.45673	&	0.17508	\\
10.61 	&	0.25761	&	0.19307	&	0.0484	&	16.91 	&	0.77005	&	0.46928	&	0.18698	\\
10.65 	&	0.25706	&	0.19185	&	0.04891	&	17.30 	&	0.78505	&	0.47031	&	0.19654	\\
10.69 	&	0.25643	&	0.19063	&	0.04935	&	17.68 	&	0.78314	&	0.46111	&	0.20238	\\
10.71 	&	0.25608	&	0.19002	&	0.04954	&	18.07 	&	0.7663	&	0.44414	&	0.20396	\\
10.75 	&	0.25532	&	0.1888	&	0.04989	&	18.45 	&	0.7386	&	0.42244	&	0.20168	\\
10.79 	&	0.25448	&	0.18755	&	0.0502	&	18.84 	&	0.70516	&	0.39907	&	0.19665	\\
10.81 	&	0.25402	&	0.1869	&	0.05034	&	19.23 	&	0.67113	&	0.37657	&	0.19045	\\
10.85 	&	0.25298	&	0.18554	&	0.05059	&	19.61 	&	0.64066	&	0.35669	&	0.18459	\\
10.87 	&	0.25233	&	0.18473	&	0.0507	&	20.00 	&	0.6166	&	0.34047	&	0.18021	\\
10.89 	&	0.25335	&	0.18555	&	0.05084	&		&		&		&		\\
\hline\hline
\end{tabular}
\end{table*}

\section{\label{sec:ion}Ionization Cross Section}
\begin{figure}[!tbp] 
\includegraphics [width=9cm]{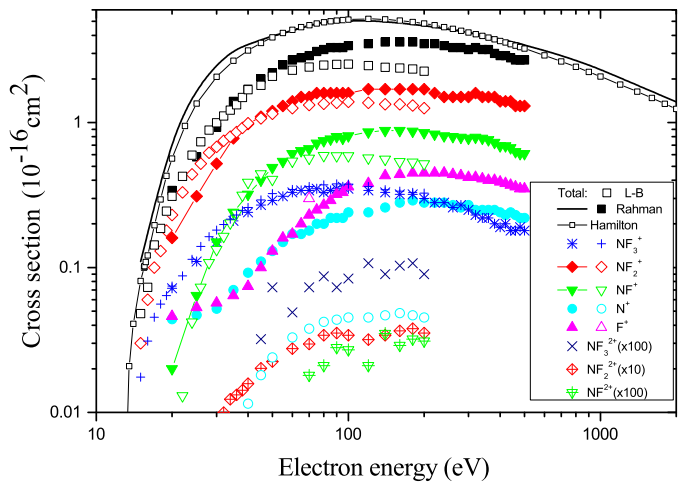}                                
\caption{\label{fig:ion01}Comparison of the compilation\cite{Landolt2000nf3} of earlier experiments with the recent measurements by Rahman {\it et al.}\cite{rahman2012electron} Closed symbols (squares, stars, diamonds, inverted triangles, circles, triangles) are Rahman {\it et al.}’s data (for total ionization, and production of NF$_3^+$, NF$_2^+$, NF$^+$, N$^+$ and F$^+$ ions, respectively).  Open symbols for NF$_3^+$, NF$_2^+$, NF$^+$ are from the Landolt-Börnstein compilation \cite{Landolt2000nf3}; an open circle and an open triangle are upper limits for N$^+$ and F$^+$ production, respectively, from Tarnovsky {\it et al.}’s experiment \cite{tarnovsky1994electron}. Crossed symbols (x, circles, inverted triangles) are L-B recommended values for doubly charged ions (NF$_3^{2+}$, NF$_2^{2+}$, NF$^+$, respectively) – the data are based on measurements by Haaland {\it et al.}\cite{haaland2001} (note the multiplying factors in figure). Thick black line is the total ionization in the complex-potential optical model by Rahman {\it et al.} \cite{rahman2012electron}, thin black line with squares is the total ionization in BEB model by Hamilton{\it et al.} \cite{hamilton2017nf3}}
\end{figure}

NF$_3$ ionization was measured in several experiments and calculated in Binary Encounter Bethe (BEB) model developed by Kim and Rudd.\cite{kim1994nf3} Generally, the agreement between experiments and the theory is rather poor: experimental data are lower than values calculated.
 Recommended values from the Landoldt-Börnstein review \cite{Landolt2000nf3} were obtained as averages of Tarnovsky {\it et al.} \cite{Tarnovsky1994nf3} and Haaland {\it et al.} \cite{haaland2001}. Total and partial ionization (into NF$_3^+$, NF$_2^+$, NF$^+$, N$^+$, NF$_3^{2+}$, NF$_2^{2+}$, NF$^{2+}$) cross sections were compiled between 14-200 eV. No data was reported of F$^+$ due to a serious disagreement between the two experiments. (Note that the figure in L-B is miscalled).  
In Fig. \ref{fig:ion01} we compare the recommended values from L-B review that resumed earlier experiments, with the recent measurements of Rahman {\it et al.}\cite{rahman2012electron}

Tarnovsky {\it et al.}\cite{tarnovsky1994electron} measured total and partial
cross sections in two laboratories (using a magnetic selector and a fast-beam
method). The agreement for the NF$_3^+$ parent ionization from the two
laboratories is within 8\%. Partial cross sections for NF$_2^+$, NF$^+$, F$^+$
ions were measured by the fast ion beam method. An upper limit for the formation
of N$^+$ was also determined. Total declared uncertainties on cross sections
were $\pm$20\%. 

Haaland {\it et al.}\cite{haaland2001} used a modified Fourier-transform mass
spectrometry: ions were confined radially by a high (2T) magnetic field and
axially by an electrostatic (1-2 V) potential. In this method no ions are
actually collected but, instead, their electromagnetic influence on the antenna
is recorded. Cross sections were normalized to Ar ionization cross sections of
Wetzel \cite{wetzel1987absolute}; the uncertainty of this normalization is
$\pm$12\% \cite{haaland2001} and the declared total uncertainty $\pm$16\%. Data
for all partial processes, including double ionizations (NF$_3^{2+}$,
NF$_2^{2+}$, NF$^{2+}$) were reported up to 200 eV. 

Rahman {\it et al.}\cite{rahman2012electron} measured total and partial (but
only for single ionization) cross sections up to 500 eV. They used a
time-of-flight spectrometer with a 30 cm-long free-flight path for ions. They
normalized relative data to the Ar$^+$ ionization cross section at 100 eV of
Krishnakumar and Srivastava \cite{krishnakumar1988ionisation}. Declared total
uncertainty was $\pm$15\%.

The best agreement (within some 15\% up to 100 eV) between the three experiments
\cite{haaland2001,tarnovsky1994electron,rahman2012electron}  is seen for the
NF$_3^+$ parent ion; at higher energies data of Rahman {\it et
al.}\cite{rahman2012electron} and Tarnovsky {\it et al.}
\cite{tarnovsky1994electron} still agree within 10\% while results of Haaland
{\it et al.} are somewhat (30\% at 200 eV) higher. For NF$_2^+$ the measurements
of Rahman {\it et al.} agree very well (within 10\%) with those by Haaland {\it
et al.} but those of Tarnovsky {\it et al.} are by 30\% lower at 200 eV.  In
turn, for NF$^+$ ion the data of Haaland are somewhat lower (20\% at 100 eV)
than the two other sets considered here. Note from Fig. \ref{fig:ion01} that the
experiment of Rahman {\it et al.}’s tends to produce higher cross sections for
NF$_2^+$ and NF$^+$ ions than the recommended data from the
Landolt-Börstein\cite{Landolt2000nf3} review; the same holds for the total
ionization.   

For light ions (N$^+$ and F$^+$) the results of Haaland {\it et
al.}\cite{haaland2001} are systematically lower (by a factor of about 10 for
F$^+$ and 50 for N$^+$) than the data by Rahman {\it et
al.}\cite{rahman2012electron}. The upper limits for N$^+$ and F$^+$ at 100 eV
given by Tarnovsky {\it et al.}\cite{tarnovsky1994electron} ($0.1\times
10^{-16}$ and $0.3\times 10^{-16}$ cm$^2$, respectively) roughly agree with
measurements by Rahman {\it et al.}, see Fig. \ref{fig:ion01}. 
Some possible systematic errors are related to the experimental methods used. In
Haaland  {\it et al.}’s \cite{haaland2001} experiment this is the indirect
measurements of the signal from ions, clearly making uncertain the determination
of partial cross sections for light ions like N$^+$ and F$^+$ (i.e. lighter than
Ar$^+$ used for normalization). On the other hand, this is the only experiment
sufficiently sensitive to detect doubly charged molecular fragments,
NF$_3^{2+}$, NF$_2^{2+}$, and NF$^{2+}$ with cross sections of the order
$10^{-20}-10^{-19}$~cm$^2$, see table \ref{tab:ion02}.

Another question is the dependence of the experimental sensitivity on the collision energy. The ion optics performs some focusing and the efficiency of ion collection can depend on their post-collisional velocities. As it was discussed for a long time in CF$_4$, see \cite{bruce1993partial}, the problem is particularly difficult when ions are formed with high velocities, and through different fragmentation channels. For NF$_3$ Tarnovsky {\it et al.}\cite{tarnovsky1994electron} observed that NF$_2^+$ ions are formed with little excess kinetic energy for impact energies near the threshold but NF$^+$ ions appear with a broad distribution of excess kinetic energy, ranging from zero to about 4 eV.
  
In time-of-flight method some metastable ions can decay before reaching detector. A rough evaluation of flight times for heavier ions give values in the microseconds range – long enough for some fragmentation to occur. This, tentatively, would explain lower values of the NF$_2^+$ cross section up to about 40 eV in Rahman {\it et al.}’s experiment as compared to the Landolt-Börnstein values, see Fig.\ref{fig:ion01}. In turn, cross sections for N$^+$ and F$^+$ show some unusual enhancement in this energy region, what would indicate some in-flight decay of heavier (NF$_2^+$, NF$^+$) ions, see Fig.\ref{fig:ion01}. 

\begin{figure}[!hbp]
\includegraphics [width=9cm]{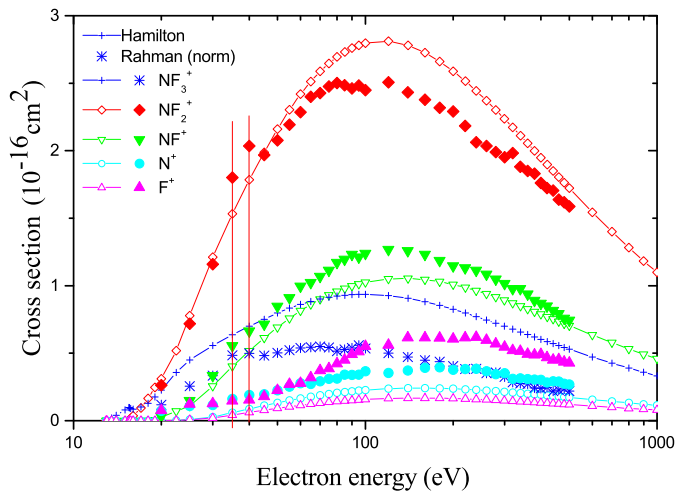}                                
\caption{\label{fig:ion02} Comparison of BEB-like partial cross sections by Hamilton {\it et al.}\cite{hamilton2017nf3} with Rahman {\it et al.} measurements \cite{rahman2012electron}. The latter partial cross sections (the sum of them) have been normalized, at each energy, to the BEB \cite{hamilton2017nf3} total ionization cross section. Vertical lines (34.6 eV and 40.4 eV) show thresholds for (F$^+$ + NF$^+$ + F) and (F$^+$ + N$^+$ +2F) channels.  }
\end{figure}
\subsection{\label{subsec:the} BEB model}
The BEB model was employed by Huo {\it et al.} \cite{huo2002nf3}, Haaland {\it et
al.} \cite{haaland2001} and Szmytkowski {\it et al.} \cite{szmytkowski2004nf3} 
to calculate the ionization cross section of NF$_3$ --
all giving the maximum for the total ionization cross section of about
$4.8\times 10^{-16}$ cm$^2$. This agrees also with the “rule-of-thumb”  noticed
recently for the CH$_4$, CH$_3$F,~...~,~CF$_4$ series \cite{Karwasz2014nf3} that 
maximum in total ionization cross section (in $10^{-16}$
cm$^2$) can be estimated as 4/3$\alpha$,
 where the dipole polarizability $\alpha$ of the molecule is expressed in
10$^{30}$ m$^3$ units. Using for NF$_3$ the dipole polarizability of $3.62\times
10^{-30}$ m$^3$ from Ref.\cite{crc2003nf3} one gets a maximum of the total
ionization cross section of $4.84\times 10^{-16}$~cm$^2$. 
Recently, Hamilton {\it et al.}\cite{hamilton2017nf3} calculated BEB ionization
cross sections using Dunning’s augmented Gaussian-type orbitals (aug-cc-pVTZ
GTO) and obtained a somewhat higher total cross section maximum
(5.19$\times 10^{-16}$ cm$^2$). Subsequently, they applied the same BEB-like
analytical expression to derive partial ionization cross sections, adapting
appropriate threshold energies. Relative amplitudes were deduced from
NIST(National Institute of Standards and Technology)\cite{nist2015nf3} electron-impact ionization mass spectrum at 100 eV.
Results for the total cross section are shown in Fig. \ref{fig:ion01} and for
partial cross sections in Fig. \ref{fig:ion02}. We adopt also these cross
sections as recommended values, see table. \ref{tab:ion01}.
We note also, that BEB total cross section agrees very well with the optical
complex-potential calculation by Rahman {\it et al.} \cite{rahman2012electron},
see Fig. \ref{fig:ion01}. 

Fig.\ref{fig:ion02} compares BEB \cite{hamilton2017nf3} partial cross sections 
 with the normalized partial cross sections of Rahman {\it et al.}. To do
this normalization, the sum of experimental partial cross sections at every
energy have been normalized to the BEB value; subsequently at this energy all
partial cross sections have been multiplied by the factor obtained. The
normalization factors range from 2.3 at 25 eV to 1.22 at 500 eV.     

This comparison allows one to distinguish difference between the expected partial
cross sections and those actually measured in the time-of-flight experiment. The
NF$_2^+$ experimental signal in the energy range above 50 eV is systematically
lower than the BEB model; the same holds for the NF$_3^+$ ion which was measured
(in all three \cite{haaland2001,tarnovsky1994electron,rahman2012electron}
experiments) to be roughly half as  abundant as the BEB \cite{hamilton2017nf3} values
(that were obtained, we recall, via NIST mass spectrum). Figure~\ref{fig:ion02}
visualizes that the deficits in the NF$_3^+$ and NF$_2^+$ experimental
abundances are compensated by higher values of F$^+$, NF$^+$ and N$^+$ signals.
This once again indicates some complex mechanisms of the dissociative
ionization.   
\begin{table}[!hbp] 
\caption{\label{tab:ion02} Ionization cross sections for formation of doubly charged ions, from Landolt-Börnstein review \cite{Landolt2000nf3} (based on Haaland {\it et al.}\cite{haaland2001}) data. Cross sections are in 10$^{-18}$ cm$^2$ units. The overall uncertainty is $\pm$50\%}
\begin{tabular}{lccc}
\hline
\hline
Energy	&	NF$_3^{2+}$	 	&	NF$_2^{2+}$ &	NF$^{2+}$ \\
(eV)	&	($10^{-18}$ cm$^2$)&	($10^{-18}$ cm$^2$)&	($10^{-18}$ cm$^2$)\\
\hline
30	&	-	&	0.0589	&	-	\\
32	&	-	&	0.0998	&	-	\\
34	&	-	&	0.123 	&	-	\\
36	&	-	&	0.132 	&	-	\\
38	&	-	&	0.142 	&	-	\\
40	&	-	&	0.158 	&	-	\\
45	&	0.032 	&	0.201 	&	-	\\
50	&	0.073 	&	0.224 	&	-	\\
60	&	0.049 	&	0.276 	&	0.000 	\\
70	&	0.073 	&	0.296 	&	0.018 	\\
80	&	0.087 	&	0.341 	&	0.021 	\\
90	&	0.073 	&	0.353 	&	0.028 	\\
100	&	0.084 	&	0.341 	&	0.027 	\\
120	&	0.107 	&	0.318 	&	0.021 	\\
140	&	0.090 	&	0.341 	&	0.035 	\\
160	&	0.103 	&	0.366 	&	0.029 	\\
180	&	0.107 	&	0.379 	&	0.032 	\\
200	&	0.090 	&	0.353 	&	0.031 	\\
\hline\hline
\end{tabular}
\end{table}
\begin{table*}[!hbp]
\caption{\label{tab:ion03} Threshold energy values (in eV) for various fragments observed and their comparison with earlier measurements. Observed thresholds are shown against the estimated ones, wherever possible.}
\begin{tabular}{lccccc}
\hline
\hline
Ions	&	Channel&	$\Delta H^o$	& Rahman {\it et al.}\cite{rahman2012electron}&Reese {\it et al.}\cite{reese1956nf3} & Tarnovsky {\it et al.}\cite{tarnovsky1994electron}	\\
\hline
NF$_3^+$	& NF$_3 \rightarrow $NF$_3^+$   & 13.5 &13.5 $\pm$ 0.6&13.2 $\pm$0.2& 13.6 $\pm$0.2\\
NF$_2^+$ & NF$_3 \rightarrow $NF$_2^+$	+F	&	14.0 &14.5 $\pm$0.6	 &14.2 $\pm$0.3  & 14.5 $\pm$0.4\\
  &  $\rightarrow $NF$_2^+$	+F$^-$	&	10.4 & - & - & -\\
  & -	&	- & 16 $\pm$0.6 & - & -\\
NF$^+$ & NF$_3 \rightarrow $NF$^+$	+2F	&	17.45 &17.5 $\pm$0.7	 &17.9 $\pm$0.3  & 17.6 $\pm$0.4\\ 
 & $\rightarrow $NF$^+$	+F$_2^-$	&	14.45&-	 &-  & -\\  
 & $\rightarrow $NF$^+$	+F+ F$^-$	&	13.85&-	 &-  & -\\  
 & $\rightarrow $NF$^+$	+F$^+$	+F 	&	34.6&-	 &-  & -\\  
 & - 	&	-&20$\pm$0.7 &-  & -\\  
F$^+$ & NF$_3 \rightarrow$F$^+$ + NF$_2^-$   & 18.8 &19 $\pm$ 1&-& -\\
 & $\rightarrow $NF + F$^+$ + F  & 22.0 &- &-& -\\
 & $\rightarrow$ N + 2F + F$^+$  & 25.85 &25$\pm$ 1 &25$\pm$ 1& -\\
 & $\rightarrow$ F$^+$ + NF$^+$ +F  &34.6 &33$\pm$ 1&-& 36\\
& $\rightarrow$ F$^+$ + N$^+$ +2F  &40.4 &- &-& -\\
N$^+$ & NF$_3 \rightarrow $N$^+$ + 3F   & 23.0 &22$\pm$ 1&22.2 $\pm$0.2& -\\
 & $\rightarrow$ N$^+$ + F + F$_2^-$   & 20.0 &-&-& -\\
 & $\rightarrow$ N$^+$ + F$^-$ + F$_2$   & 19.4 &-&-& -\\
\hline\hline
\end{tabular}
\end{table*}
\begin{table*}[!hbp] 
\caption{\label{tab:ion01} Recommended ionization cross sections for NF$_3$: BEB values from ref. \cite{hamilton2017nf3}, see text).  Energies are in eV, cross sections are in 10$^{-16}$ cm$^2$ units. The overall uncertainty is $\pm$10\% for total cross sections and $\pm$20\% for partial ones.  }
\begin{tabular}{lcccccclcccccc}
\hline
\hline

Energy	&	N$^+$&	F$^+$	&	NF$^+$ &	NF$_2^+$ &	NF$_3^+$	&	Total	&Energy	&	N$^+$&	F$^+$	&	NF$^+$ &	NF$_2^+$ &	NF$_3^+$	&	Total\\
\hline

14	&	-	&	-	&	-	&	-	&	0.04	&	0.04	&	200	&	0.24	&	0.16	&	1.01	&	2.59	&	0.83	&	4.83	\\

15	&	-	&	-	&	-	&	-	&	0.08	&	0.08	&	220	&	0.23	&	0.16	&	0.99	&	2.52	&	0.80	&	4.69	\\

16	&	-	&	-	&	-	&	0.04	&	0.09	&	0.12	& 240	&	0.23	&	0.16	&	0.96	&	2.44	&	0.77	&	4.56	\\
17	&	-	&	-	&	-	&	0.10	&	0.10	&	0.20	& 260	&	0.22	&	0.16	&	0.94	&	2.37	&	0.75	&	4.44	\\
18	&	-	&	-	&	-	&	0.17	&	0.13	&	0.30	&	
280	&	0.22	&	0.15	&	0.92	&	2.30	&	0.72	&	4.31    \\
19	&	-	&	-	&	0.01	&	0.25	&	0.18	&	0.43	&
300	&	0.21	&	0.15	&	0.89	&	2.24	&	0.70	&	4.19	\\

20	&	-	&		&	0.03	&	0.31	&	0.23	&	0.56	&	

320	&	0.21	&	0.15	&	0.87	&	2.17	&	0.68	&	4.08	\\

25	&	0.01	&	0.01	&	0.11	&	0.78	&	0.45	&	1.35	&
340	&	0.21	&	0.14	&	0.85	&	2.11	&	0.66	&	3.96	\\

30	&	0.03	&	0.02	&	0.26	&	1.21	&	0.56	&	2.07	&
360	&	0.20	&	0.14	&	0.83	&	2.06	&	0.64	&	3.86	\\

35	&	0.06	&	0.04	&	0.40	&	1.53	&	0.64	&	2.67	&	

380	&	0.20	&	0.14	&	0.81	&	2.00	&	0.62	&	3.76	\\

40	&	0.09	&	0.06	&	0.52	&	1.79	&	0.70	&	3.15	&
400	&	0.19	&	0.13	&	0.79	&	1.95	&	0.60	&	3.66	\\

45	&	0.11	&	0.08	&	0.61	&	1.99	&	0.75	&	3.54	&
420	&	0.19	&	0.13	&	0.77	&	1.90	&	0.59	&	3.57	\\

50	&	0.13	&	0.09	&	0.69	&	2.16	&	0.80	&	3.88	&
440	&	0.18	&	0.13	&	0.75	&	1.85	&	0.57	&	3.48	\\

55	&	0.15	&	0.1	&	0.76	&	2.30	&	0.83	&	4.15	&	
460	&	0.18	&	0.12	&	0.74	&	1.81	&	0.56	&	3.40	\\

60	&	0.17	&	0.11	&	0.81	&	2.42	&	0.86	&	4.38	&
480	&	0.18	&	0.12	&	0.72	&	1.76	&	0.54	&	3.32	\\
65	&	0.18	&	0.12	&	0.86	&	2.51	&	0.88	&	4.56	&	
500	&	0.17	&	0.12	&	0.70	&	1.72	&	0.53	&	3.25	\\

70	&	0.19	&	0.13	&	0.90	&	2.59	&	0.90	&	4.71	&	
600	&	0.16	&	0.11	&	0.64	&	1.54	&	0.47	&	2.92	\\

75	&	0.20	&	0.14	&	0.93	&	2.65	&	0.91	&	4.83	&	700	&	0.14	&	0.10	&	0.58	&	1.40	&	0.43	&	2.65	\\
80	&	0.21	&	0.14	&	0.96	&	2.69	&	0.92	&	4.93	&
800	&	0.13	&	0.09	&	0.53	&	1.28	&	0.39	&	2.43	\\

85	&	0.21	&	0.15	&	0.98	&	2.73	&	0.93	&	5.00	&
900	&	0.12	&	0.09	&	0.49	&	1.18	&	0.36	&	2.24	\\	
90	&	0.22	&	0.15	&	1.00	&	2.76	&	0.93	&	5.06	&
1000	&	0.11	&	0.08	&	0.46	&	1.10	&	0.33	&	2.08	\\
95	&	0.22	&	0.16	&	1.01	&	2.78	&	0.94	&	5.11	&	

1200	&	0.10	&	0.07	&	0.40	&	0.96	&	0.29	&	1.83	\\

100	&	0.23	&	0.16	&	1.03	&	2.80	&	0.94	&	5.14	&	
1400	&	0.09	&	0.06	&	0.36	&	0.86	&	0.26	&	1.63	\\

120	&	0.24	&	0.16	&	1.05	&	2.81	&	0.93	&	5.19	&	1600	&	0.08	&	0.06	&	0.33	&	0.78	&	0.23	&	1.48	\\
140	&	0.24	&	0.17	&	1.05	&	2.78	&	0.91	&	5.15	&	1800	&	0.08	&	0.05	&	0.30	&	0.71	&	0.21	&	1.35	\\
	
160	&	0.24	&	0.17	&	1.05	&	2.73	&	0.88	&	5.06	&
2000	&	0.07	&	0.05	&	0.28	&	0.66	&	0.20	&	1.25	\\

180	&	0.24	&	0.17	&	1.03	&	2.66	&	0.85	&	4.95	\\

\hline\hline
\end{tabular}
\end{table*}
\subsection{\label{subsec:DISS} Dissociative ionization channels}
Ionization threshold were thoroughly reported by Rahman {\it et al.} 
\cite{rahman2012electron} and Tarnovsky {\it et al.} 
\cite{tarnovsky1994electron}. The threshold for NF$_3^+$ parent ion is 
13.6$\pm$0.2 eV \cite{tarnovsky1994electron}.  The NF$_2^+$ ion shows, according 
to Rahman {\it et al.}, two onsets, closely spaced (14.5 and 16 eV); note that 
only one value (14.5$\pm$0.4 eV \cite{tarnovsky1994electron}) was reported by 
other experiments. see table \ref{tab:ion03}.
Three onsets were identified by Tarnovsky {\it et al.} for the NF$^+$ ion: 
(NF$^+$+ 2F) at 17.6 eV, (F$^+$+NF$^+$+F) at 36.5 eV (reported by Rahman {\it et 
al.}  \cite{rahman2012electron} in the F$^+$ production channel) and another, 
not identified, at 21.8 eV, see table 3. Rahman {\it et al.} reported a 
threshold of 19$\pm$1 eV for F$^+$ production and assigned it to the 
NF$_3\to$F$^+$ + NF$_2^-$ process. Note that this is the only reporting on the dipolar dissociation in NF$_3$.
\subsection{\label{subsec:rec} Recommended values}

Taking into account the significant differences between experiments (and their different sensitivities), 
we decided to recommend total and partial ionization cross sections from the recent BEB analysis by Hamilton {\it et al.}\cite{hamilton2017nf3}, detailed results of which are given in table \ref{tab:ion01}. We stress that such a choice in not free of some coarse assumptions:  the validity of BEB model for partial cross sections and correct procedures in measurements of partial ions in NIST experiment. We are also aware that our explanations for the differences between the most recent \cite{rahman2012electron} and earlier experiments\cite{haaland2001,tarnovsky1994electron} are only tentative. Therefore, we refer the reader to data by Rahman {\it et al.}\cite{rahman2012electron} (table 2 in their paper) as a complementary choice. In view of importance of NF$_3$ for plasma processes in semiconductor industries, verification of cross section for total and partial ionization is urgent; new theories would be also welcome. The estimated uncertainty on presently recommended values is $+$10\% $-$20\% for the total ionization and $+$20\% $-$30\% for partial cross sections.

\section{\label{sec:att} Electron Attachment (DEA) Cross Section}

\begin{figure}[!tbp]
\includegraphics [width=7cm]{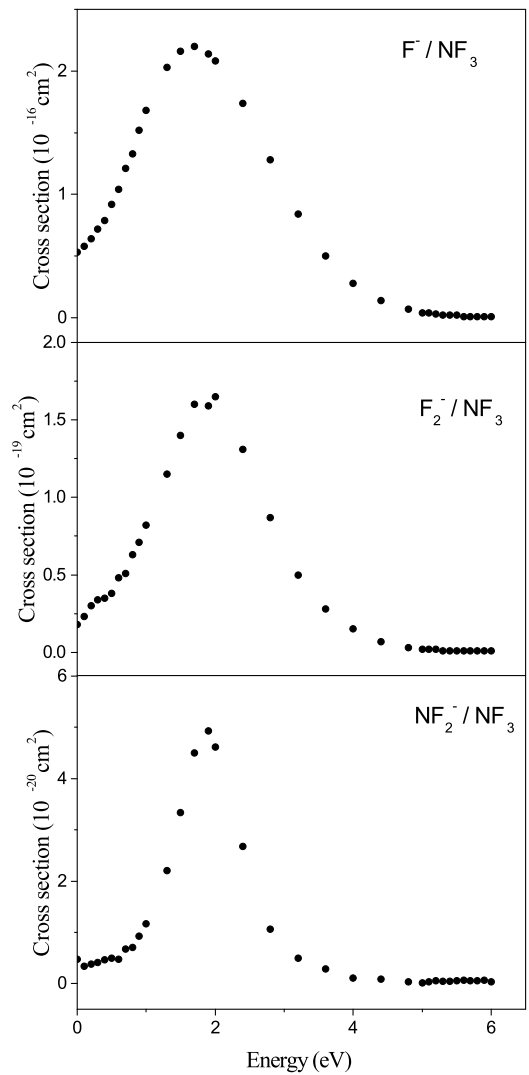}                                
\caption{\label{fig:dea}. Recommended cross sections for the formation of F$^-$, F$_2^-$ and NF$_2^-$ from NF$_3$. Note the different scales in cross sections\cite{nandi2001nf3}.}
\end{figure}

There are only two published reports, relevant to this evaluation purpose, on
the absolute measurements of the dissociative electron attachment (DEA) cross
sections for NF$_3$: Harland and Franklin \cite{Harland1974nf3} and Nandi {\it
et al.} \cite{nandi2001nf3} Other than these, Chantry reported the DEA cross
sections for NF$_3$  at a conference and the results were also contained in a book
\cite{chantry1979dissociative} but were not published in a journal, and
therefore, this will not be discussed any further here. 
There are two experimental determinations of the electron attachment coefficient in swarms,\cite{nbh79,lakdawala1980nf3},
but the two results are not consistent with each other and it is unclear which is more reliable.
Harland and Franklin \cite{Harland1974nf3} employed a linear TOF mass spectrometer
to measure translational energies of negative ions formed by dissociative
resonance capture processes from NF$_3$. Nandi {\it et al.} \cite{nandi2001nf3}
have pointed out that mass to charge ratio analysis becomes imperative for the
measurement of partial cross sections when more than one type of ions are
produced and then it is necessary that the extraction, mass analysis, and the
detection procedures for these ions are carried out without discriminating
against their initial kinetic energies, angular distributions or their mass to
charge ratios. 
These necessitated them to use a crossed beam geometry and an efficient solution
to these problems was to use a segmented time-of-flight (TOF) mass spectrometer
along with the pulsed-electron-beam and pulsed-ion-extraction techniques and the
relative flow technique.
The results of Harland and Franklin \cite{Harland1974nf3} and Nandi {\it et al.}
\cite{nandi2001nf3} agree well with each other in the positions of the resonance
peaks, but the magnitude show the differences as nearly big as a factor of four.
For example, the cross sections of the formation of 
F$^-$ are 0.6 $ \times 10^{-16}$ cm$^2$ for Harland and Franklin and 2.2 
$\times 10^{-16}$ cm$^2$ for Nandi {\it et al.} In both experiments, F$^-$ is the
most dominant ion from the DEA process with very small intensities of F$_2^-$
and NF$_2^-$. For F$_2^-$, the cross section of Nandi {\it et al.} is smaller
than that obtained by  and Franklin. For NF$_2^-$, the cross section of Nandi
{\it et al.} is larger than the corresponding data of Harland and Franklin
within a factor of 2. For all the ions, there is a finite cross section even at
zero energy. The electron beam has a halfwidth of 0.5 eV. It is possible that
the high energy tail of the electron energy distribution is giving rise to the
finite cross section at zero energy for F$_2^-$ and NF$_2^-$
\cite{nandi2001nf3}. The high resolution measurements of Ruckhaberle {\it et
al.} \cite{ruckhaberle1997nf3} showed finite cross section for F$^-$ at zero
energy, whereas both F$_2^-$ and NF$_2^-$ appear only above 1 eV
\cite{nandi2001nf3}. Considering the fact that Nandi {\it et al.} have made more complete measurements of the experimental parameters, we recommend their cross sections for the dissociative electron attachment process of NF$_3$. Complete numerical values of the recommended cross sections are presented in Table
\ref{tab:dea} and Fig. \ref{fig:dea}. Nandi {\it et al.} estimated the
uncertainty to be about 15\%.

\begin{table}[!tbp]
\caption{\label{tab:dea} Recommended dissociative attachment cross sections for the formation of F$^-$, F$_2^-$ and NF$_2^-$ from NF$_3$\cite{nandi2001nf3}.The uncertainties are estimated to be about 15 \%.}
\begin{tabular}{lccc}
\hline
\hline
Energy	&	$\sigma$(F$^-$)	&	$\sigma$(F$_2^-$)	&	$\sigma$(NF$_2^-$)	\\
(eV)	&	($10^{-16}$cm$^2$)&	($10^{-19}$cm$^2$)	&	($10^{-20}$cm$^2$)\\
\hline
\hline
0.0 	&	0.53 	&	0.18 	&	0.48 	\\
0.1 	&	0.58 	&	0.23 	&	0.34 	\\
0.2 	&	0.64 	&	0.30 	&	0.38 	\\
0.3 	&	0.72 	&	0.34 	&	0.41 	\\
0.4 	&	0.79 	&	0.35 	&	0.46 	\\
0.5 	&	0.92 	&	0.38 	&	0.50 	\\
0.6 	&	1.04 	&	0.48 	&	0.47 	\\
0.7 	&	1.21 	&	0.51 	&	0.67 	\\
0.8 	&	1.33 	&	0.63 	&	0.71 	\\
0.9 	&	1.52 	&	0.71 	&	0.93 	\\
1.0 	&	1.68 	&	0.82 	&	1.17 	\\
1.3 	&	2.03 	&	1.15 	&	2.20 	\\
1.5 	&	2.16 	&	1.40 	&	3.34 	\\
1.7 	&	2.20 	&	1.60 	&	4.50 	\\
1.9 	&	2.14 	&	1.59 	&	4.93 	\\
2.0 	&	2.08 	&	1.65 	&	4.62 	\\
2.4 	&	1.74 	&	1.31 	&	2.68 	\\
2.8 	&	1.28 	&	0.87 	&	1.06 	\\
3.2 	&	0.84 	&	0.50 	&	0.50 	\\
3.6 	&	0.50 	&	0.28 	&	0.29 	\\
4.0 	&	0.28 	&	0.15 	&	0.11 	\\
4.4 	&	0.14 	&	0.07 	&	0.09 	\\
4.8 	&	0.07 	&	0.03 	&	0.03 	\\
5.0 	&	0.04 	&	0.02 	&	0.01 	\\
5.1 	&	0.04 	&	0.02 	&	0.03 	\\
5.2 	&	0.03 	&	0.02 	&	0.06 	\\
5.3 	&	0.02 	&	0.01 	&	0.05 	\\
5.4 	&	0.02 	&	0.01 	&	0.05 	\\
5.5 	&	0.02 	&	0.01 	&	0.06 	\\
5.6 	&	0.01 	&	0.01 	&	0.07 	\\
5.7 	&	0.01 	&	0.01 	&	0.06 	\\
5.8 	&	0.01 	&	0.01 	&	0.06 	\\
5.9 	&	0.01 	&	0.01 	&	0.07 	\\
6.0 	&	0.01 	&	0.01 	&	0.03 	\\
\hline\hline
\end{tabular}
\end{table}

\section{\label{sec:sum}Summary and Future work}

\begin{figure} [!tbp]                                          
\includegraphics [width=8cm]{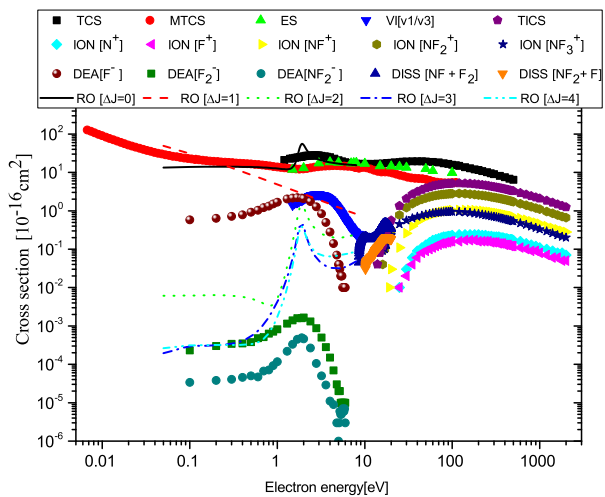}                                  
\caption{\label{fig:sum}The summary of cross section for electron collisions with $\rm{NF_3}$.
TCS - total scattering, ES - elastic scattering, MT - momentum transfer,
ION - partial ionization, TICS - total ionization, VI - vibrational excitation, RO - rotational excitation, DEA - dissociative electron attachment, DISS - Neutral dissociation cross section }
\end{figure}

We  present a systematic review of 
the published cross sections for processes resulting from
electron collisions with NF$_3$ up to the end of 2016. Both measurements and
theoretical predictions are considered, although priority is given
to high quality measurements with published uncertainties where available.
The summary of cross section for electron collisions with NF$_3$ is given in Fig. \ref{fig:sum}.
There is considerable variation
in the reliability of the available data. For the total cross section, the momentum transfer
cross section and the ionization cross section, it is possible to recommend values over an
extended energy range with small uncertainties, typically 10 to 15\%. The situation is significantly worse for other processes. For electron impact 
rotational excitation we rely on predictions from {\it ab initial} calculations, but these calculations are far from being complete. The experimental work on this process would be welcome. There is one
direct experimental measurements of electron impact vibrational excitation cross sections. Theoretical treatments of this process is possible and should be performed by theorists. Some
new, reliable beam measurements of this process would be very helpful. Electron impact
dissociation  is an important process but the available measurements are inconsistent
with each other and we are unable to recommend a good set of data for this process. A new study
on the problem is needed. Finally there are two data available for the dissociative
electron attachment process. Here we recommend using the most recent experimental data and are able to provide estimated uncertainty to be about 15\%.

This evaluation is one  in series of systematic evaluations \cite{song2015cross,song2017HCCH} of electron collision processes for key molecular targets. Other evaluations will appear in future papers.

\section{Acknowledgments}
This work was partially supported by the National Research Council of Science \& Technology (NST) grant by the Korea government (MSIP) (No. PCS-17-05-NFRI). H. C. acknowledges the support from Chungnam National University in 2016. VK acknowledges partial support from the National Science Foundation, Grant No PHY-15-06391. GK acknowledges partial support from Grant 2014/15/D/ST2/02358 of National Science Center in Poland.
\section{\label{sec:ref}References}

%
\end{document}